\documentclass[12pt,letterpaper]{article}
\usepackage[includeheadfoot, 
            marginratio={1:1,2:3}, 
            width=412pt, 
            height=688pt,]{geometry}

\usepackage{amsmath}
\usepackage{amsfonts}
\usepackage{amssymb}
\usepackage{graphicx}
\usepackage{subfigure}

\allowdisplaybreaks[2]
\numberwithin{equation}{section}

\newcommand{\eq}[1]{\begin{equation}
                     \begin{split} #1 \end{split}
                     \end{equation}}

\newcommand{\V}{\mathcal{V}}      
\newcommand{\K}{\mathcal{K}}

\newcommand{\ov}[1]{\overline{#1}}               
\newcommand{\be}{\begin{equation}}
\newcommand{\ee}{\end{equation}}
\newcommand{\ba}{\begin{array}}
\newcommand{\ea}{\end{array}}
\newcommand{\bea}{\begin{eqnarray}}
\newcommand{\eea}{\end{eqnarray}}
\def\IP{\relax{\rm I\kern-.18em P}}

\begin{document}


\vspace*{-1.5cm}
\begin{flushright}
  {\small
  MPP-2007-169 \\
  }
\end{flushright}

\vspace{2.3cm}
\begin{center}
  {\LARGE Moduli Stabilisation versus Chirality for \\[0.2cm]
       MSSM like Type IIB Orientifolds 
  }
\end{center}

\vspace{0.65cm}
\begin{center}
  Ralph Blumenhagen, 
  Sebastian Moster, Erik Plauschinn \\
\end{center}

\vspace{0.1cm}
\begin{center}
  Max-Planck-Institut f\"ur Physik, F\"ohringer Ring 6, \\
  80805 M\"unchen, Germany  \\
\end{center}

\vspace{-0.1cm}
\begin{center}
  \tt{
  blumenha, moster, plausch @mppmu.mpg.de \\
  }
\end{center}

\vspace{1.5cm}
\begin{abstract}

We investigate the general question of implementing a chiral
MSSM like D-brane sector in Type IIB orientifold models 
with complete moduli stabilisation via F-terms induced by 
fluxes and space-time instantons, respectively gaugino condensates.
The prototype examples are the KKLT and the so-called large volume 
compactifications. 
We show that the 
ansatz of first stabilising all moduli via F-terms and then introducing
the Standard Model module is misleading, as a chiral sector
notoriously influences the structure of non-perturbative effects
and induces a D-term potential.
Focusing for concreteness on the large volume scenario, 
we work out the geometry of  the  swiss-cheese type Calabi-Yau manifold
$\IP_{[1,3,3,3,5]}[15]_{(3,75)}$ and 
analyse whether controllable and
phenomenologically  acceptable K\"ahler moduli stabilisation
can occur by the combination of  F- and D-terms.

\end{abstract}

\thispagestyle{empty}
\clearpage
\tableofcontents


\section{Introduction}

It is the main goal of string phenomenology to
find realistic string models which are predictive
(in the weak sense\footnote{We are not discussing
here problems related to the plethora of string models
(the string landscape) and its consequences for  
the predictivity of string theory.}) and  which, for a fixed background,
fix  dynamically all
low energy parameters.
Indeed, it would be an important advance and a proof of principle 
to find a globally consistent
string compactification which contains and combines all the various
mechanisms of moduli stabilisation and D-brane model building
techniques leading to
the MSSM localised on some (intersecting) D-branes 
 and leading to a predictive framework for cosmology. 
This means it would allow us to precisely compute all MSSM and 
cosmological parameters from the underlying dynamically stabilised 
string model.

In recent years some progress has been made towards
actually achieving this goal in that
new ways of fixing moduli in string compactifications have
been found (see \cite{Uranga:2003pz,MarchesanoBuznego:2003hp,Lust:2004ks,Blumenhagen:2005mu,Grana:2005jc,Douglas:2006es,Blumenhagen:2006ci,Denef:2007pq} for reviews).
The most often discussed cases are Type IIB orientifolds,
where a combination of background fluxes \cite{Giddings:2001yu} and
non-perturbative terms were argued to allow to fix
all closed string moduli \cite{Kachru:2003aw}. Here the complex structure
moduli and the dilaton are stabilised by three-form fluxes
and the K\"ahler moduli by non-perturbative terms
arising from D3-brane instantons and/or gaugino
condensation on D7-branes 
(for a systematic analysis see 
for \linebreak instance \cite{Denef:2004dm,Robbins:2004hx,Denef:2005mm,Lust:2005dy,Lust:2006zg}).\footnote{In the remainder
of this paper we will mainly consider D3-brane instantons.
However, let us already mention that the results we
 obtain  carry over to gaugino condensation on D7-branes.} 

Quite remarkably, generalising the KKLT scenario,
phenomenologically appealing
models have been found, for which in a non-supersymmetric minimum
of the scalar potential the overall volume ${\cal V}$ of
the Calabi-Yau manifold is fixed  
at a very large value like for instance ${\cal V}\simeq 10^{15}$, 
where $\alpha'$-corrections
to the K\"ahler potential compete with non-perturbative contributions
to the superpotential \cite{Balasubramanian:2005zx}. The K\"ahler moduli,
on whose associated cycles the euclidean D3-brane instantons are wrapped,
are fixed by non-perturbative contributions to the superpotential
at small volume, $\tau\simeq \log ({\cal V})$.

These models have been called large volume
compactifications and many of their low energy features 
have been worked out, including a computation
of supersymmetry breaking soft terms \cite{Conlon:2005ki,Conlon:2006wz} and its
collider signatures \cite{Conlon:2007xv}. 
We think it is fair to say that
such an investigation has set a new standard for
doing string phenomenology.

Let us point out that the main implicit assumption
made in the KKLT and large volume scenarios  was that it is a valid
procedure 
to split the construction of such string models into two steps.
The first one is to fix
all moduli by a combination of fluxes and non-perturbative effects.
After that has been achieved, the second step is to  introduce
the {\it module} of the MSSM on some intersecting respectively
magnetised D7-branes. 
On a phenomenological level this might be a fair
attempt. It is the aim of this paper to 
investigate more closely, whether the conditions
appearing  in string theory justify such a procedure.
For concreteness and because of their attractive phenomenological properties, 
in this paper we will mainly discuss the large volume scenario but would like 
to stress that  all the structure and constraints we find directly carry over to other
constructions including the KKLT scenario \cite{Kachru:2003aw}.

Essentially generalising the arguments of 
\cite{Choi:2005ge,Achucarro:2006zf,Lebedev:2006qq,Villadoro:2006ia},
in this paper we would like to emphasise that there is 
a  fundamental problem when
combining a chiral MSSM like module  with the moduli stabilisation
module. 
After reviewing the main ingredients of the large
volume scenario in section 2, we will point out two general
features which a chiral D7-brane sector introduces:
\begin{itemize}
\setlength{\itemsep}{2pt}
\item{
It leads to the generation of
a D-term potential, which appears at lower
order in the $1/{\cal V}$ expansion of the scalar potential.
Therefore, 
there is the danger of destabilising the large volume
minimum.
}
\item On the intersection
of the D7-branes with the E3-brane instantons extra
charged {\it chiral} fermionic zero modes can appear,
spoiling the generation of an uncharged superpotential.

\end{itemize}

We will argue that, as studied for instance
in  \cite{Achucarro:2006zf,Lebedev:2006qq,Cremades:2007ig}, a 
non-perturbative superpotential including   matter fields 
does not resolve the second problem, as in our bottom-up approach
we do not want to give VEVs to
MSSM matter fields. This is in contrast to  \cite{Achucarro:2006zf,Parameswaran:2006jh,Lebedev:2006qq,Cremades:2007ig}, where 
these matter field superpotentials were used in the
hidden sector for uplifting the AdS minimum to de Sitter.
There, no direct phenomenological constraints arise 
for the VEVs of the hidden sector matter fields.  

As we will show, due to the chirality of the D7-brane sector, 
in our MSSM case the D7-branes and the E3-branes
should better wrap in some sense orthogonal four-cycles.
This means that not all the sizes of the 
D7-branes are  fixable  by instanton induced F-terms. 
However, it is precisely the sizes of these  cycles on which
some of the low-energy MSSM parameters depend.
Therefore, by this effect we seem to  loose part of the very predictive power
of the models \cite{Conlon:2005ki,Conlon:2006wz,Conlon:2007xv}.

The combination of both aspects mentioned in the previous
paragraph provides a natural solution to the
problem of fixing all K\"ahler moduli for an MSSM like model. 
Non-perturbative effects
fix part of the K\"ahler moduli except some of the  ones controlling the size
of the MSSM branes. These latter are fixed by the vanishing
of the D-term potential. Since it is a D-term, there could
also be charged matter contributions. Again, by requiring
not to break the MSSM gauge symmetry already at the high scale they
should better have vanishing vacuum expectation values.
Note that, for the uplifting physics in the hidden sector
this argument does not apply.

Furthermore, we find that in the case of 
 multiple contributions to the instanton
generated superpotential, there are additional terms potentially
destabilising the large volume minimum found in 
\cite{Balasubramanian:2005zx,Conlon:2005ki} for  
the single instanton case. However, requiring 
that the four-cycles the instantons wrap do not intersect
leads to the familiar form of the scalar potential \cite{Balasubramanian:2005zx,Conlon:2005ki}.
We also allow for more general rigid four-cycles, homologically 
described by linear combinations of the basic blow-up
cycles which can be understood as rigid, singular configurations
of the basic ones. In this case, the F-terms also take a slightly
different form than described in \cite{Balasubramanian:2005zx,Conlon:2005ki}.

In the second part of this paper, we will investigate some of these aspects 
for the swiss cheese type Calabi-Yau manifold defined via
the resolution of the singular hypersurface in a weighted projective space
$\IP_{[1,3,3,3,5]}[15]_{(3,75)}$.\footnote{
In Appendix \ref{app_cy2}, we will also provide the geometric data for a second new
swiss-cheese Calabi-Yau, namely the resolution of
$\IP_{[1,1,3,10,15]}[30]_{(5,251)}$.}
Working out the toric geometry of the resolution,
we will show that this  space contains  rigid four-cycles,
on which wrapped E3-instantons have the right zero mode
structure to give a contribution to the uncharged superpotential.
Introducing  also a set of chirally intersecting and magnetised D7-branes,
we find that, for vanishing VEVs for charged matter fields, 
the combination of F- and D-terms fixes the four-cycles 
at the boundary of the K\"ahler cone, where
however the sizes of the instantons and D7-branes remain finite
and of the same scale. For non-vanishing VEVs of some of the matter
fields the situation gets even improved.


\section{Type IIB orientifolds}

The best understood moduli stabilisation techniques have
been developed for the perturbative Type IIB string, which
is mainly related to the fact that turning on background three-form
fluxes only mildly changes the background geometry by 
introducing a warp factor \cite{Giddings:2001yu}. 
As we will explain at the example of the large volume models of
\cite{Balasubramanian:2005zx,Conlon:2005ki}, 
the task of  combining moduli stabilisation with an
MSSM type gauge sector is non-trivial and more involved
than just combining these two separate modules.


\subsection{Large volume scenario}
\label{sec_lvs}

In order to explain our arguments, we will use as a prototype example
the large volume scenario (LVS) of closed string moduli stabilisation in 
Type IIB orientifold models \cite{Balasubramanian:2005zx}.
We consider Type IIB orientifolds of a Calabi-Yau manifold ${\cal X}$
with in general O7- and O3-planes. The complex structure moduli $U$ are encoded in the holomorphic three-form $\Omega_3$ of ${\cal X}$
and the axio-dilaton field  reads
\bea 
         S=e^{-\phi} + i\,C_0 \;,
\eea
where $\phi$ is the dilaton and $C_0$ is the Ramond-Ramond (R-R) zero-form. These moduli are usually fixed by the Gukov-Vafa-Witten superpotential \cite{Gukov:1999ya}
\bea 
  \label{w_GVW}
  W_{\rm{GVW}} = \int_{\mathcal{X}} G_3\wedge \Omega_3 \;,
\eea
arising from
three form flux $G_3=F_3+ iS\, H_3$ supported on the three-cycles
of the Calabi-Yau manifold. 
As for the KKLT scenario \cite{Kachru:2003aw}, it is assumed that the flux vacuum breaks
supersymmetry, where the value of the superpotential in the minimum
is denoted by $W_0$.

Concerning the stabilisation of the K\"ahler moduli, the tree-level no-scale structure of the
flux induced scalar potential is broken by both the leading
order perturbative $\alpha'$-corrections to the tree-level
K\"ahler potential and by E3-brane instanton corrections
to the superpotential. The K\"ahler potential including the 
$\alpha'$-corrections reads \cite{Becker:2002nn} 
\eq{
  \label{kahl-pot}
  K = -2\ln \biggl( \hat{\V} + \frac{{\xi}}{2\, g_s^{3/ 2}} \biggr) - 
  \ln \biggl( S + \ov S \biggr)
  - \ln \biggl( -i\int_{\mathcal{X}} \Omega\wedge\ov \Omega \biggr) 
  \;,
}
where $g_s$ denotes the string coupling and we have set the 
supergravity scale as $M_{\rm Pl}=1$. 
The string-frame volume $\V$ of the Calabi-Yau manifold is expressed in the following way
\bea
  \V = \frac{1}{3!}\int_{\mathcal{X}} J\wedge J\wedge J 
  = \frac{1}{6}\, \K_{ijk}\, t^i\, t^j\, t^k \;,
\eea
where we have expanded the K\"ahler form $J$ in some basis 
$\{\omega_i\}$ of $H^{1,1}(\mathcal{X},\mathbb{Z})$ as $J=\sum_i t^i
\omega_i$. The Einstein-frame volume appearing in \eqref{kahl-pot} is denoted by a hat. Furthermore, $\K_{ijk}$ denotes the triple intersection number 
in the chosen basis. The $\alpha'$-corrections are encoded in 
$\xi$ in terms of the Euler number $\chi$ of the internal manifold $\mathcal{X}$
\bea
  {\xi} = -\frac{\zeta(3)\, \chi\bigl(\mathcal{X}\bigr)}{2\:(2\pi)^3}\;.
\eea
In addition to the perturbative corrections, one also takes into account E3-brane instantons in order to break the no-scale structure of the scalar potential. These instantons generate terms in the superpotential of the following form \cite{Witten:1996bn}
\bea
\label{instanton}
  W_{\rm{np}}= A(S,U)\, e^{-S_{\rm{inst}}}=A(S,U)\, e^{-\sum_i a_i 
  T_i} \;.
\eea
The K\"ahler moduli $T_i$ of Type IIB orientifolds with O3- and O7-planes are a particular combination of the K\"ahler form $J$ and of the R-R four-form $C_4$ 
\bea
   T_i = e^{-\phi}\:\frac{1}{2}\:\int_{D_i} J\wedge J 
   + i \int_{D_i}C_4 
   = e^{-\phi}\: \tau_i + i\: \rho_i \;,
\eea
where $\{D_i\}\subset H_4(\mathcal{X},\mathbb{Z})$ with
$i=1,\ldots,h_{1,1}(\mathcal{X})$ forms a basis of four-cycles on the internal manifold. 
The $\tau_i$ denote the volume of $D_i$ 
and $\rho_i$ is the associated axion. For ease of notation, we
will express all geometric quantities like $\V$ and $\tau$ in
string-frame. However, in the supergravity formulas, they have to
appear in Einstein-frame which is achieved by substituting
$J\to e^{-\phi/2}J$. As already indicated above, the resulting quantities in this frame will be
denoted by a hat, i.e. $\hat\V$ and $\hat\tau$.

For manifolds where the overall volume $\V$ is controlled by one large 
four-cycle, it was shown that such compactifications admit an interesting 
minimum of the resulting scalar potential \cite{Balasubramanian:2005zx}. 
More precisely, the minimum arises at exponentially large volumes, if $\V$ 
can be written as
\bea
    {\cal V}\sim \tau_b^{\frac{3}{2}}-\sum_{s=1}^{h_{1,1}-1} 
    \tau_s^{\frac{3}{2}},
\eea
where $\tau_b$ denotes the volume of the big four-cycle and $\tau_s$ measure 
the sizes of (small) holes in this geometry. Thus, these models have 
a ``swiss-cheese'' like structure \cite{Conlon:2005ki,Conlon:2006gv,Berg:2007wt}.

The standard example primarily studied in the literature is the Calabi-Yau manifold
$\IP_{[1,1,1,6,9]}[18]$ which has a codimension three
$\mathbb{Z}_3$-singularity \cite{Denef:2004dm,Curio:2006ea,Misra:2007yu}. 
Its resolution introduces a second K\"ahler 
modulus $T_s$ so that the volume becomes
\bea
    {\cal V}={1\over 9\sqrt{2}}\left( \tau_b^{3\over 2}-\tau_s^{3\over 
    2} \right) \;.
\eea
Since the minimum of the scalar potential is expected to occur at large values of 
$\V$, the leading order instanton contribution 
$W_{\rm{np}}=A_s\, e^{-a_s T_s}$ is given by an E3-brane instanton along the 
small cycle. Furthermore, because $\V\gg1$, one can perform an expansion of 
the potential $V_F$ in powers of $1/\V$ \cite{Balasubramanian:2005zx}
\begin{align}
  V_F =&\:\: e^{K} \left( G^{a\ov b} \, D_a W \, D_{\ov b}\ov W - 3 
  \bigl| W \bigr|^2 \right) \nonumber \\
   =& \:\:
  \lambda\,{(a_sA_s)^2\, \sqrt{\hat\tau_s}\:e^{-2a_s\hat\tau_s} \over {\cal \hat V}} 
  - \mu\,{  a_s \bigl|A_sW_0\bigr|\: \hat\tau_s\,  e^{-a_s\hat\tau_s}
  \over 
  {\cal \hat V}^2} +
  \nu\,{ \xi\: \bigl|W_0\bigr|^2 \over g_s^{3/2}\: {\cal \hat V}^3} + 
  \ldots\;.
  \label{scapot}
\end{align}
Here $\lambda,\mu,\nu$ are positive numerical
constants and $a_s$ takes the value $2\pi$. The complex structure moduli $U$ and the axio-dilation are assumed 
to  be stabilised via $D_{U}W=D_SW=0$ and sub-leading powers of 
$1/{\cal \hat V}$ have been neglected. 
For $\xi>0$ this potential then has a minimum which stabilises 
${\cal \hat V}\simeq \hat\tau_b^{3/2}$ at large values and $\hat\tau_s$ at 
$\hat\tau_{s}\simeq\log({\cal \hat V})$. Introducing the mass scale
$M_{\rm Pl}$, in this scenario one obtains
the relation between the string and the Planck scale as
$M_s\simeq M_{\rm Pl}/\sqrt{\cal V}$ and the 
relation $M_{3\over 2}\simeq M_{\rm Pl}\, W_0/{\cal V}$ for the gravitino mass.

In subsequent papers \cite{Conlon:2005ki,Conlon:2006us,Conlon:2006wz,Conlon:2006wt,Conlon:2007xv}, many phenomenologically appealing features
of this scenario have been found, which at the moment
make them very attractive candidates for string phenomenology.
Moreover, it has been shown in 
\cite{Haack:2006cy,Cicoli:2007xp} that due to a second 
no-scale structure string loop corrections to the K\"ahler potential are 
sub-dominant.

\bigskip
To prepare our following discussion let us mention two important points:

\begin{itemize}

\item{In order for an E3-brane instanton to actually generate
a superpotential term like in (\ref{instanton}), the zero
mode structure must be of a special nature. For the example 
$\IP_{[1,1,1,6,9]}[18]$ it was shown that the small divisor $\tau_s$ has an
F-theory lift to a six-dimensional divisor in the Calabi-Yau 
fourfold with $\chi(D,{\cal O})=1$.
By the zero mode criterion derived in \cite{Witten:1996bn}, 
an instanton along this cycle thus contributes to the superpotential.
}

\item{In all the phenomenological analysis of the model above,
    the MSSM D7-branes were assumed to also wrap the small cycle
       $\tau_s$. Therefore, the implicit philosophy was, that
       one first freezes all closed string moduli by fluxes and
      instantons, and then adds an MSSM like D7-brane sector
      and computes the soft
      terms depending on the non-vanishing auxiliary F-fields
      via the usual supergravity formulas \cite{Brignole:1997dp}. }

\end{itemize}

This latter practice is surely justified
in a purely phenomenological approach, but, given the 
undoubted success of this scenario, we would like
to look more closely whether such a procedure is indeed
justified from the structure of string theory. 
Of course, there are apparent stringy consistency conditions, which
are not shown to be really satisfied in this concrete model, 
like for instance tadpole cancellation conditions and
the vanishing of the Freed-Witten anomalies \cite{Freed:1999vc} appearing
if both $H_3$-form flux and D-branes are present. 
If these are violated then the theory would be inconsistent
right away as in general anomalies would not be cancelled.\footnote{
In a T-dual Type IIA model, where the Standard Model moduli 
can also be fixed by fluxes, it would not only be the interplay between
intersecting D6-branes and E2-instantons but also  the generalised Freed-Witten \cite{Villadoro:2005yq,Villadoro:2006ia}
anomalies governing the coexistence  between the
chiral gauge sector and the moduli stabilisation sector.} 
What we are after, however, is more subtle
and relates to the coexistence  of a chiral D-brane sector and
a moduli freezing instanton sector.


\subsection{Orientifolds with intersecting D7-branes}

Let us now collect the general rules for computing the massless spectrum and the
tadpole cancellation conditions for Type IIB orientifolds
with O7- and O3-planes.
As before, we are compactifying the Type IIB string on a Calabi-Yau three-fold
${\cal X}$ but now we also
specify an orientifold projection. It is of the form  
$\Omega\sigma(-1)^{F_L}$ where $\Omega$ is the world-sheet parity operator, 
$\sigma$ is an involution and $F_L$ denotes the left-moving fermion number. 
Then, we introduce stacks of D7$_a$-branes wrapping four-cycles $D_a$ in the Calabi-Yau manifold and carrying 
gauge bundles $V_a$. Here we consider only orientifold projections
leaving all four-cycles invariant, i.e. 
 $h^{-}_{1,1}=0$ and $h^{+}_{1,1}=h_{1,1}$.
This implies that the orientifold projection acts as
\eq{
  \Omega_3\to - \Omega_3\;,\qquad
  D_a \to D_a\;,\qquad
  V_a \to V_a^{\vee} ,
}
where $\Omega_3$ denotes  the holomorphic three-form of the Calabi-Yau. 
The fixed point locus of the involution $\sigma$ defines a
divisor $D_{\rm{O7}}$ around which the orientifold plane is wrapped.
Note that on the O7-plane there is no gauge bundle so that 
formally we choose $V_{\rm{O7}}=\mathcal{O}$.

The chiral massless spectrum arising from open strings stretched
between two D7-branes wrapping two four-cycles
$D_a$ and $D_b$ and carrying gauge bundles $V_a$ and $V_b$
is determined by \cite{Aspinwall:2004jr,Douglas:2006xy}
\bea
  \label{intersection_iib}
  I_{ab} = \int_{D_a\cap D_b} \Bigl( c_1(V_a)-c_1(V_b) \Bigr) 
  =\int_{\cal X}
   \Bigl(c_1(V_a)-c_1(V_b) \Bigr)\wedge [D_a] \wedge [D_b] \;.
\eea
Here, the two-forms $[D_{a,b}]$ denote the Poincar\'{e} duals to the 
four-cycles $D_{a,b}$ and $c_1(V_{a,b})$ denote the first Chern classes of 
$V_{a,b}$. 
The rules for computing the chiral spectrum are summarised in table \ref{tab_chir_spec} where a prime denotes the orientifold image.
\begin{table}[tb]
\centering
  \renewcommand{\arraystretch}{1.1}
\begin{tabular}{c||c}
  Representation & Multiplicity \\ \hline\hline \\[-5mm]
  $(\overline{N}_a,N_b)$ & $I_{ab}$ \\
  $(N_a,N_b)$ & $I_{a'b}$ \\
  $A_a$ & $\frac{1}{2}( I_{a'a}+2I_{\rm{O7}a})$ \\
  $S_a$ & $\frac{1}{2}( I_{a'a}-2I_{\rm{O7}a})$ 
\end{tabular}
\caption{Chiral spectrum for intersecting D7-branes.\label{tab_chir_spec}}
\end{table}

Having a chiral spectrum implies that one has to worry about anomalies. 
However, satisfying the tadpole cancellation condition for the D7-branes ensures that the 
spectrum is free of non-abelian gauge anomalies. For the present case it reads
\eq{
\label{tadseven}
  \sum_a  N_a\, D_a = 4\, D_{\rm{O7}} \;,
}
where the sum is over all D7$_a$-branes. Note that we have presented
the tadpole constraint  on the orientifold quotient. 
In the ambient Calabi-Yau it is multiplied by a factor of two. 
In addition, there is the D3-brane tadpole which, again on the
quotient, takes the following form
\eq{
  \label{tapole_d3}
  N_{\rm{D3}}+N_{\rm{flux}}- \sum_a N_a \int_{D_a}\!\! 
  \mbox{ch}_2\bigl( V_a \bigr)={N_{\rm O3}\over 4}
  + \sum_a \frac{N_a}{24} \int_{D_a}\!\!  
    \mbox{c}_2\bigl( T_{D_a} \bigr)  
  +\frac{1}{12} \int_{D_{\rm{O7}}}\!\! 
    \mbox{c}_2\bigl( T_{\rm{O7}} \bigr) \;,
}
where $T_D$ denotes the tangential bundle  of the divisor $D$ and $c_2$ 
stands for the second Chern class while ch$_2$ denotes the second Chern character.
Note that for a smooth divisor $D$ the integral of the second Chern
class over $D$ is just the Euler-characteristic $\chi(D)$.

In the F-theory lift of such a model to a  Calabi-Yau fourfold ${\cal Y}$, 
the right hand side of equation \eqref{tapole_d3} is equal to $\chi(\mathcal{Y})/24$ \cite{Sethi:1996es}.
For the simple solution of (\ref{tadseven})
with four D7-branes with trivial line bundle placed right on top of the
O7-plane we have
\bea
\label{drauf}
  N_{\rm{D3}}+N_{\rm{flux}}={N_{O3}\over 4}+{\chi(D_{\rm O7})\over 4}\;.
\eea
The gauge group in this case is $SO(8)$.
For this special solution, the  Calabi-Yau four-fold 
is given by the $\mathbb{Z}_2$ orbifold
${\cal Y}=({\cal X}\times \mathbb{T}^2)/\mathbb{Z}_2$ where the $\mathbb{Z}_2$ acts as 
the holomorphic involution $\sigma$ on the CY three-fold ${\cal X}$ and 
on the torus $\mathbb{T}^2$ as $z\to -z$. 
If we blow-up the
$\mathbb{Z}_2$ singularities by gluing in $\IP^1$s and take
the four fixed points on $T^2$ into account, the
Euler-characteristic of ${\cal Y}$ is computed as
\begin{align}
   \chi\bigl(\mathcal{Y}\bigr)&={1\over 2}\biggl( \chi\bigl({\cal X}\times \mathbb{T}^2\bigr) 
   - 4\, \chi\bigl(D_{\rm O7}\bigr)
   -4 N_{\rm O3} \biggr) + 4\, \chi\bigl(\IP^1\bigr)\chi\bigl(D_{\rm O7}\bigr)
   +  4\, \chi\bigl(\IP^1\bigr) N_{\rm O3} \nonumber \\
   &= 24\, \left(  {N_{O3}\over 4}+{\chi(D_{\rm O7})\over 4}
                          \right)\;.
\end{align}
For other non-trivial and in particular chiral solutions
to the tadpole cancellation conditions, the F-theory four-fold
is  not explicitly known.\footnote{Note that it is somewhat misleading
to do Type IIB model building with the fourfold base $F_{11}$, 
as here implicitly the four-cycles wrapped by  the D7-branes have already
been fixed and the only freedom is to turn on gauge bundles
on them.}


\section{Instantons  and chirality}

In this section, we are going to investigate the aforementioned
interplay between a 
chiral theory realised by intersecting and magnetised D7-branes 
and the E3-brane instantons. 
More specifically, we assume that some version of the MSSM can be described by a
configuration of D7-branes wrapping four-cycles in the Calabi-Yau
manifold.


\subsection{The chiral D7-brane sector}
\label{sec_chiral_d7}

The formula for the chiral spectrum between two D7-branes 
(\ref{intersection_iib})
implies that in order to obtain chirality, it is necessary that
at least one of the  D7-branes carries a non-trivial $U(N)$ gauge bundle.
For our purposes, it is not crucial to have a complete MSSM sector,
but we will just take one of the main features of the Standard Model, namely its 
chirality,  and assume the minimal chiral configuration.
We consider $K$ stacks of $N_a$ D7-branes wrapping the cycle $D_{a}$ with 
vector bundle $V_a$. However, in order to avoid stability issues
of higher rank vector bundles and vector bundle
moduli, from now on we just choose line bundles ${{\cal L}_a}$ on the 
D7-branes.

For such chiral intersecting D-brane models, it is known
that generically they contain anomalous $U(1)$ gauge symmetries.
For D7-branes, these anomalies are cancelled by the four-dimensional axions
\bea
           \rho_{a} =  \int_{D_{a}} C_4
\eea
arising from the dimensional reduction of the R-R four-form along the
four-cycle $D_{a}$. Indeed, the Chern-Simons action for a D7-brane on a 
four-cycle $D_a$ contains terms of the form
\bea
   S_{\rm{CS}} \sim  \int_{\mathbb{R}^{1,3}\times D_{a}} C_4\wedge 
   F\wedge F \;,
\eea
which give rise to the following Green-Schwarz couplings. First, there is the 
mass-term for the gauge field obtained by choosing two legs of $C_4$ along 
$D_{a}$ and $F$ to be the curvature
of the internal line bundle ${\cal L}$. Second, the $\rho-A^2$ vertex arises
from choosing all four legs of $C_4$ along $D_{a}$.
Such a gauging of the axionic shift symmetry leads to a 
Fayet-Iliopoulos term 
for a $U(1)$, which in our case turns out to be
\bea
   \label{FI_parameter}
   \xi_{a}= {1\over {\cal \hat V}} 
     \int_{\cal X}    c_1\bigl( {\cal L}_{a} \bigr)
     \wedge \bigl[D_{a}\bigr]\wedge \hat J\; .
\eea
Therefore, a chiral D7-brane sector necessarily gives rise to
a D-term potential $V_D$ of the following form
\bea
\label{dterm}
  V_D = \sum_{a=1}^K \, \frac{1}{\mbox{Re}\,(f_{a})} \left( \sum_i  Q^{(a)}_i \bigl| 
  \phi_{i} \bigr|^2 - 
  \;\xi_{a} \right)^2 \;,
\eea
where $M_{\rm Pl}=1$ and $Q^{(a)}_i$ are the $U(1)_{a}$ charges of the canonically
normalised matter fields $\phi_i$. 
Furthermore, $\mbox{Re}\,(f_{a})$ denotes the real part of the gauge kinetic function for 
the corresponding D-brane. It is effectively the DBI action of a 
supersymmetric E3-brane instanton along the cycle $D_a$ and reads
\eq{
  \mbox{Re}\,(f_{a})= e^{-\phi}\:\frac{1}{2}\:\int_{D_a} J\wedge J
  - e^{-\phi} \int_{D_a} \mbox{ch}_2\bigl(B+\mathcal{L}_a \bigr) 
  = \hat\tau_{a} - \mbox{Re}\bigl( S \bigr)\: c_{a} \;.
}
Here, $c_{a}$ denotes the integrated second Chern character of 
$B+\mathcal{L}_a$ on the respective D7-brane and $\hat\tau_a$ is the (Einstein-frame) volume of $D_a$.

\bigskip
Note that this D-term is generically only of order ${\cal V}^{-2}$ in the 
volume expansion \eqref{scapot}
so that an additional (natural) D-term supersymmetry breaking  
destabilises the 
large volume minimum found at order ${\cal V}^{-3}$. 
Therefore, for preserving the large volume minimum we  will
require that the 
D-term  vanishes, i.e. $V_D=0$. The other option is to allow for
significant fine tuning and use this D-term in a hidden sector
for up-lifting the AdS minimum
to a small and positive vacuum energy 
\cite{Burgess:2003ic,Dudas:2005vv,Villadoro:2005yq,Achucarro:2006zf}.


\subsection{E3-brane instantons}

We are now going to investigate E3-brane instanton effects in more detail. In particular, for 
the instantons to generate a contribution to the superpotential, the zero 
mode structure has to be of a certain type.

Note that for orientifold models where the D7-branes do 
not lie right on top of the O7-planes, i.e. for generally intersecting D-branes
with non-trivial gauge bundles, the F-theory lift is not known and the 
computation of $\chi(D,{\cal O})$ for the uplifted E3-brane divisor cannot be 
performed. It is therefore much more convenient to perform the zero mode analysis
directly in the Type IIB orientifold model, where we can rely
on recent work on space-time instanton effects in D-brane models
\cite{Billo:2002hm,Blumenhagen:2006xt,Ibanez:2006da,Bianchi:2007fx,Akerblom:2007uc,Grimm:2007xm,Blumenhagen:2007bn,Cvetic:2007qj}.
We will also comment on the case that we freeze the K\"ahler moduli
by  gaugino condensates on a stack of $N_c$ D7-branes wrapping
a four-cycle $D_G$.

\begin{itemize}

\item For contributing to the superpotential, a single (isolated) instanton must wrap a four-cycle invariant 
under the orientifold projection and must carry an $O(1)$ gauge symmetry 
\cite{Argurio:2007vqa,Bianchi:2007wy,Ibanez:2007rs}. 
In the case of $h_{1,1}^-(\mathcal{X})=0$, this implies that  the instanton carries a trivial gauge bundle.

\item Next, we have to worry about deformation zero modes of the E3-instanton.
These are clearly absent, if the E3-brane wrapping the four-dimensional 
divisor $D$ does 
not have any further moduli. That is, there are no Wilson lines counted by 
$H^1(D,{\cal O})$ or transverse deformations counted by $H^2(D,{\cal O})$. 
If this sufficient condition
is not satisfied, then fluxes or curvature on the moduli space
might soak up some of the zero modes, but a more careful analysis
is necessary \cite{Blumenhagen:2007bn,GarciaEtxebarria:2007zv}.
Similarly, for gaugino condensation many adjoint matter fields counted by
$H^{i}(D_G,{\cal O})$ with $i=1,2$ 
spoil asymptotic freedom of the gauge theory on the $D7_G$-branes.

\item If, as in our case, there are additional space-time filling D7-branes 
present, there can appear extra charged fermionic zero modes from the 
intersection of the E3-instanton and the D7-branes \cite{Blumenhagen:2006xt}. 
The chiral index of these fermionic zero modes is
\bea
  \label{zeroferm}
   Z_{a}= N_a\, \int_{D_{a}\cap D_{\rm{E3}}} c_1
   \bigl({\cal L}_a\bigr)
   =N_a\,\int_{\cal X} c_1\bigl({\cal L}_a\bigr) \wedge 
   \bigl[D_{a}\bigr] \wedge \bigl[D_{\rm{E3}}\bigr] \;.
\eea
In order to soak up these  additional fermionic zero 
modes, one has to pull down charged matter fields in 
the instanton computation. The pure exponential term as in (\ref{instanton}) 
is then multiplied by products of charged matter superfields 
$\Phi_i$ as  \cite{Blumenhagen:2006xt}
\bea
  \label{w_stringy}
  W_{\rm{string}} \sim \biggl[\prod_i \Phi_i\biggr]\:
   e^{-\mathcal{S}_{\rm{inst}}}.
\eea
Note, that such instantons are not gauge instantons
and therefore often called  stringy or exotic  instantons.

\item For the special case when the E3-instanton lies right on top of the 
D7-branes\footnote{Note that in \cite{Ibanez:2007tu,Petersson:2007sc} it was
  shown that for an instanton 
on top of a single D-brane also a superpotential of the form 
\eqref{w_stringy} can be generated.}, it is possible to have non-trivial gauge bundles on the instanton. 
It can then be regarded  as a gauge instanton from the perspective of the 
D7-brane gauge theory and additional bosonic and non-chiral fermionic 
zero modes arise parametrising the ADHM instanton super moduli space 
\cite{Billo:2002hm,Akerblom:2006hx,Argurio:2007vqa,Bianchi:2007wy}.
The effect of such instantons is of the same nature as gaugino
condensates for the gauge theory on stacks of D7$_G$-branes, so
that we can discuss them together. In order to soak
up the ADHM zero modes one needs extra non-chiral (with respect to
the $U(N_G)$ gauge group) matter zero modes
from the intersection of the E3-instanton with the other
D7-branes \cite{Haack:2006cy,Akerblom:2006hx}. 
If we end up with an $SU(N_c)$ gauge group
with effectively $N_f$ flavours, then for $N_f<N_c$ the contribution
to the superpotential is
\bea
  W_{\rm{gauge}} \sim {1\over 
   \det_{ff'}\left[\widetilde{\Phi}_{\!f}^{\:\:c}\, 
   \Phi_{c\,f'}\right]}\: 
   e^{-\mathcal{S}_{\rm{inst}}}  \;.
\eea
In writing this, it  is assumed that we are on the Higgs branch, where
the determinant is non-vanishing and so the flavour
gauge group is completely broken. Such a configuration is not part of the MSSM and therefore the instanton respectively the 
$D_G$ branes should better not have any intersection with
the D-branes supporting the MSSM.
\end{itemize}


\subsection{Moduli stabilisation for chiral  models}
\label{sec_moduli_stab}

We will now argue that given the structure and constraints 
from the previous discussion, for chiral orientifolds not all K\"ahler moduli 
can be frozen by instantons. In particular, some of the moduli controlling 
the size of the chiral D7-brane sector are left unfixed by the E3-brane 
instantons.

\bigskip
Let us first summarise the possible matter fields which can be present 
in the configurations we are considering.
\begin{itemize}

\item We assume that the chiral MSSM like matter fields, 
denoted as $\Phi_{\rm{SM}}$,  are part of the
chiral matter spectrum arising on a set of  intersecting D7-branes
carrying initial gauge group $G=\prod_{a=1}^K U(N_a)$. 
Typical  examples discussed in the literature are $G=U(5)\times U(1)$,
$G=U(4)\times U(2)\times U(2)$ 
or $G=U(3)\times U(2)\times U(1)\times U(1)$.

\item There can also be  additional (chiral) fields, 
which also arise from the same set of intersecting D7-branes
leading to so-called exotic matter fields.
There can exist exotic matter fields
transforming in non-trivial representation of the non-abelian
part of the MSSM gauge group. 
These are denoted as $\Phi_{\rm{exo}}$.

\item However, since in D-brane models we genuinely have these
extra $U(1)$ gauge factors, there might be fields
which are not charged under the MSSM gauge group $SU(3)\times SU(2)\times
U(1)_Y$ but carry non-trivial  charges with respect
to $U(1)$s orthogonal to $U(1)_Y$.
These we denote as $\Phi_{\rm{abel}}$.

\item In addition there can in principle
be further hidden sector matter fields $\Phi_{\rm{H}}$, whose D-terms and 
F-terms however do not mix with the Standard Model ones. Therefore, we will 
not focus on those in the following. However, this sector might be important
for the eventual uplift of the AdS minimum to de Sitter with small
cosmological constant.

\end{itemize}

Consider now an  E3-instanton wrapping a four-cycle which gives rise to extra 
Standard Model
charged zero modes. These can either be chiral fermionic zero modes
coming from stringy instantons or non-chiral zero modes from gauge instantons.
To soak up all these zero modes,  the superpotential coupling must contain
products of the Standard Model Matter fields $\Phi_{\rm{SM}}$ and, since they 
appear on  the same D7-branes, also products of the additional fields $\Phi_{\rm{exo}}$ 
and $\Phi_{\rm{abel}}$
\bea
  \label{suppot_sm}
  W \sim  \prod_i \Phi_{\rm{SM}}^{(i)}\, \prod_j   
  \Phi_{\rm{exo}}^{(j)}\, 
  \prod_k \Phi_{\rm{abel}}^{(k)}\,\,\,  e^{-T_{E3}} \;.
\eea
Note that for gauge instantons or gaugino condensates there will
be determinants of the matter fields in the denominator. Furthermore,
in the equation above $T_{\rm{E3}}=\sum_i m^iT_i$ denotes the K\"ahler modulus corresponding 
to the instanton on the cycle $D_{\rm{E3}}=\sum_i m^i D_i$. 

The important 
point is now that, for phenomenological reasons,
at this high scale we do not want to break
the MSSM gauge symmetry by giving VEVs to these fields.
If we  allow for VEVs of charged matter fields, the
D-term potential (\ref{dterm}) generates a mass of the generic
order $M_{\rm matter}=M_{\rm Pl}/\sqrt{\cal V}=M_s$ for them, i.e.
the matter fields become very heavy. 
The MSSM gauge symmetry breaking and mass generation
should occur as usual at the low scale in the process
of supersymmetry breaking.
Therefore, we are only interested in vacua with 
$\langle \Phi_{\rm{SM}}\rangle=\langle \Phi_{\rm{exo}}\rangle=0$, so that
effectively the contribution of such an instanton to the superpotential 
 vanishes 
and the F-term potential $V_F$ does not depend explicitly on $T_{\rm{E3}}$. 
What could be possible in principle is to allow VEVs for GUT Higgs fields.

Of course this argumentation is not really satisfying as in a fully realistic 
moduli stabilisation scenario, we also would like
to have these charged matter fields dynamically stabilised.
But our point of view is, that it is very likely that in a given concrete model the 
four contributions: \footnote{See for instance \cite{GarciadelMoral:2005js,Maillard:2007pq} for a recent
discussion of matter fields moduli stabilisation.}
\begin{itemize}
\item the soft supersymmetry breaking mass terms $V_{\rm soft}= m^2 \Phi^2_{\rm SM}$,
\item the perturbative and instanton induced superpotential  contributions of
  the form  $W=\prod \Phi_{\rm SM}$,
\item the D-terms  and
\item the generic absence of gauge instantons or gaugino condensates
      for MSSM fields, i.e. terms like
  $W_{\rm{gauge}} \sim {1\over \det\left[ \Phi_{\rm SM} \right]}\: 
  e^{-\mathcal{S}_{\rm{inst}}}$
\end{itemize}
suffice to freeze
to MSSM matter fields at $\langle \Phi_{\rm{SM}}\rangle=\langle
\Phi_{\rm{exo}}\rangle=0$.
If such a mechanism is indeed at work, then, since they appear
in the same open string sector, also the fields $\Phi_{\rm abel}$
are likely to be frozen at vanishing VEVs. However, just from phenomenology 
these VEVs could be non-vanishing, a fact to be kept  in mind when  we will mainly discuss
the case $\langle
\Phi_{\rm{abel}}\rangle=0$.

Therefore, if we want to fix the size of the four-cycle the E3-instanton is 
wrapping, it should not have any zero modes charged under the Standard Model 
gauge symmetry. Recall that we derived the analogous condition also
for moduli freezing via gaugino condensates on a stack of D7-branes
wrapping a four-cycle $D_G$. There too, $D_G$ should not have any
charged matter fields from intersections with branes supporting the
MSSM. 

Recalling then equation \eqref{zeroferm}, we have to satisfy 
the necessary condition
\bea
  \label{zerofermzero}
  N_a\int_{\cal X}
  c_1\bigl({\cal L}_a\bigr) 
  \wedge \bigl[D_{a}\bigr] \wedge \bigl[D_{\rm{E3}}\bigr]=0\;,
\eea
for Standard Model branes wrapping the divisor $D_a$ with line bundle 
${\cal L}_a$. 
Furthermore, not only the chiral instanton zero modes have to be absent but 
 also those which are vector-like. For determining them one has
 to compute the cohomology  classes
\bea
  H^{i}\left( D_a\cap D_{\rm{E3}} \, ,\, {\cal L}_a\otimes 
  {\cal K}^{1\over 2}_{D_a}\otimes
  {\cal K}^{1\over 2}_{D_{\rm{E3}}}\right)
  \qquad  {\rm for}\quad i=0,1\; ,
\eea
where ${\cal K}_{D}$ denotes the canonical line bundle of the divisor
$D\subset {\cal X}$. If these cohomology classes are non-trivial, extra pairs of 
instanton zero modes are present and the resulting term in the superpotential 
will be of the form \eqref{suppot_sm}. However, in this paper
we will mainly be concerned with chiral zero modes and generically
do not explicitly determine the vector-like ones. But one has
to keep in mind that they might be present and one has to worry
about soaking them up.

Coming back to equation \eqref{zerofermzero}, we can expand the Poincar\'{e} 
dual of the instanton cycle $[D_{\rm{E3}}]$ in a basis $\{\omega_i\}$ of
 two-forms in $H^{1,1}({\cal X})$
\bea
  \bigl[D_{\rm{E3}}\bigr] = \sum_i  m^i\,\, \omega_i \;.
\eea
Then, we define the following matrix
\bea
  \label{matrix1}
  \mathcal{M}_{a,i}=  \int_{\cal X}
  c_1\bigl(\mathcal{L}_a\bigr) \wedge \bigl[D_{a}\bigr] 
  \wedge \omega_i\;,
\eea             
with $i=1,\ldots, h_{1,1}({\cal X})$ and $a=1,\ldots, K$ where $K$ is the
number of MSSM supporting D7-branes carrying $U(N)$ gauge symmetry . 
To not over-constrain the system, we can assume that 
$K\le h_{1,1}$  and so the maximal number of linear independent 
E3-brane instantons ${\cal N}_{\rm{E3}}$
one is allowed to introduce is given by the kernel of the matrix $\mathcal{M}_{a,i}$.

\bigskip
Since the kernel of the matrix \eqref{matrix1} is not equal to 
$h_{1,1}({\cal X})$ because of the chirality of the MSSM, it is clear that not all K\"ahler
moduli can be stabilised by E3-brane instantons. 
But let us expand the K\"ahler form $J$ in the basis $\{\omega_i\}$ as 
$J=\sum_i  t^i\,\,\omega_i$. Recalling then equation \eqref{FI_parameter}, 
we find that the Fayet-Iliopoulos parameter can be expressed as
\bea
   \label{FI_parameter_2}
   \xi_{a}= {1\over {\cal \hat V}} 
   \int_{\cal X}    c_1\bigl( {\cal L}_{a} \bigr)
   \wedge \bigl[D_{a}\bigr]\wedge \hat J
   = \frac{1}{\hat \V}\sum_i \mathcal{M}_{a,i}\; \hat t^i \;,
\eea
so that the K\"ahler moduli will also appear in the D-terms. The vanishing of 
the D-terms then provides additional restrictions on the $t^i$. The number of moduli 
fixed through these equations is given by the rank of the matrix $\mathcal{M}_{a,i}$ which
satisfies ${\rm rk}(\mathcal{M})\ge K_{\rm anom}$ where $K_{\rm anom}$ denotes
the number of anomalous $U(1)$ gauge factors supported on the MSSM branes. 
To be more precise, the K\"ahler moduli counted by the defect of $\mathcal{M}_{a,i}$ are fixed by the D-term. These are orthogonal to the ones possibly fixed by E3-instantons and are in the kernel of $\mathcal{M}_{a,i}$.
Since the MSSM matter spectrum is chiral, it is 
clear from the definition of $\mathcal{M}_{a,i}$ that there
must be at least one anomalous $U(1)$ gauge factor. 

To summarise:
at most $h_{1,1}-K_{\rm anom}$ K\"ahler moduli can be
fixed by E3-instantons whereas 
for the remaining moduli, which control
the size of the D7-branes supporting the MSSM sector, there appears a D-term potential.
For not destabilising  the large volume minimum due to the ${1/{\cal V}^2}$ 
factor in front, this D-term has to vanish.
Therefore, despite our initial concern, with sufficient rigid instantons being present
in a  model, we have enough constraints to fix all K\"ahler moduli. 
If we cannot fix all remaining K\"ahler moduli via instantons and D-terms,
there also exist the possibility that they are  frozen  similar to ${\cal V}$
by perturbative corrections to the F-term scalar potential. 
Arguments have been given that this should occur for the
QCD axion \cite{Conlon:2006tq}. \footnote{We thank Joe Conlon for bringing
this to our attention.}

Clearly, the general arguments presented above need to be investigated 
more carefully for each model, this is however beyond the scope of this paper.
From now on, if not dynamically proven but at least phenomenologically
motivated, we generally assume 
\bea
\langle \Phi_{\rm{SM}}\rangle=\langle \Phi_{\rm{exo}}\rangle=\langle \Phi_{\rm{abel}}\rangle=0 \;,
\eea
so that the vanishing of the D-terms in the MSSM sector effectively implies the vanishing of the 
Fayet-Iliopoulos parameters \eqref{FI_parameter_2}.
We will mention at certain points the changes once VEVs of $\Phi_{\rm abel}$
are nonvanishingg, but as we stressed already so far we do not have a 
complete theory to dynamically freeze these moduli.


\subsection{F-term scalar potential}
\label{sec_scapot_fterm}

In the original work about the large volume scenario 
\cite{Balasubramanian:2005zx}, only the case with one E3-brane instanton 
along one small four-cycle was studied in detail. Later it was argued that 
similar results carry over to configurations where more than one four-cycle 
stays small supporting instantons \cite{Conlon:2006gv}. For our purpose it is 
useful and illustrative to start again from a general setup and perform the 
steps along the lines of \cite{Balasubramanian:2005zx}.

Similarly to section \ref{sec_lvs}, we assume that the complex structure moduli $U$ and the axio-dilaton $S$ have been fixed by fluxes via $D_UW=D_SW=0$ and
the value of the Gukov-Vafa-Witten superpotential \eqref{w_GVW} in the minimum will again be denoted by $W_0$.
For the stabilisation of the K\"ahler moduli we use the usual $\alpha'$-corrected K\"ahler potential \eqref{kahl-pot} and introduce E3-instantons.
However, we we allow for instantons wrapping general four-cycles $D_{\alpha} = M_{\alpha}^i D_i$ where $M_{\alpha}^i$ are the wrapping numbers of the instanton $\alpha$ and $\{D_i\}$ is a basis of four-cycles on $\mathcal{X}$. 
The superpotential then takes the form
\eq{
  W = W_0 + \sum_{\alpha} A_{\alpha} \: e^{-2\pi M_{\alpha}^i T_i} \;,
}
where the sum is over all contributing instantons in the large radius
limit.
Computing  
the K\"ahler metric similarly to \cite{Bobkov:2004cy}, we can write the 
scalar F-term potential as
\eq{
  \label{potential-F_00}
  V_F = e^K \Biggl(\; &
    -\frac{(2\pi)^2}{2} 
    \left( 2\hat{\V} + \hat{\xi}\right)
    \sum_{\alpha,\beta} \mbox{Vol}\bigl( D_{\alpha} \cap D_{\beta} 
    \bigr)\:
    A_{\alpha}\: \ov A_{\beta}\: e^{-2\pi M_{\alpha}^i T_i} \: 
    e^{-2\pi M_{\beta}^j \ov T_j}  \\[-2mm]
  &+\frac{(2\pi)^2}{4}\: \frac{4\hat{\V}-\hat{\xi}}      
    {\hat{\V}-\hat{\xi}}
    \sum_{\alpha,\beta} \hat\tau_{\alpha}\: \hat\tau_{\beta}\:
    A_{\alpha}\: \ov A_{\beta}\: e^{-2\pi M_{\alpha}^i T_i} \: 
    e^{-2\pi M_{\beta}^j \ov T_j}  \\
  &+\frac{2\pi }{2}\: \frac{4 \hat{\V}^2+ \hat\V\hat\xi+4\hat\xi^2}
    {\bigl( 2\hat\V+ \hat\xi \bigr) \bigl( \hat\V-\hat{\xi}\bigr)}
    \sum_{\alpha} \hat{\tau}_{\alpha}\: 
    \Bigl( A_{\alpha} \: e^{-2\pi M_{\alpha}^i T_i} \ov W 
    + \ov A_{\alpha}\: e^{-2\pi M_{\alpha}^i \ov T_i} W \Bigr) \\
  &+3\:\hat\xi\: \frac{ \hat\V^2+7 \hat\V\hat\xi + \hat\xi^2}
    {\bigl( 2\hat\V+ \hat\xi\bigr)^2\bigl( \hat\V-\hat\xi \bigr)}\:
    \bigl| W \bigr|^2
    \quad\Biggr) \;.
}
Here we have used $\hat\V$ and $\hat{\tau}_{\alpha}$ to respectively denote in Einstein-frame the volume of the Calabi-Yau manifold and the volume of the four-cycle wrapped by the instanton $\alpha$. Furthermore, to simplify the formulas we used 
\eq{
  \mbox{Vol}\bigl( D_{\alpha} \cap D_{\beta} \bigr)
  = M_{\alpha}^i\: M_{\beta}^j\: \K_{ijk} \hat{t}^k 
}
for the volume of the intersection of two four-cycles $D_{\alpha}$ and $D_{\beta}$ (in Einstein-frame) and we have defined $\hat\xi=\xi/g_s^{3/2}$.

Let us now perform the large volume expansion of $V_F$. Note that in this limit the second term in \eqref{potential-F_00} is sub-leading. Keeping also only the leading term $W_0$ in the superpotential, we find
up to an overall constant
\eq{
  \label{potential-F_01}
  V_F \simeq &
    -\frac{(2\pi)^2}{\hat\V}\:
    \sum_{\alpha,\beta} \mbox{Vol}\bigl( D_{\alpha} \cap D_{\beta} 
    \bigr)\:
    A_{\alpha}\: \ov A_{\beta}\: e^{-2\pi M_{\alpha}^i T_i} \: 
    e^{-2\pi M_{\beta}^j \ov T_j}  \\
  &+\frac{2\pi}{\hat\V^2}\:
     \sum_{\alpha} \hat{\tau}_{\alpha} \:
     \Bigl( A_{\alpha} \: e^{-2\pi M_{\alpha}^i T_i} \:\ov W_0 
    + \ov A_{\alpha}\: e^{-2\pi M_{\alpha}^i \ov T_i}\: W_0 \Bigr) 
    +\frac{3}{4}\: \frac{\hat\xi}{\hat\V^3}\: \bigl| W_0 \bigr|^2 \;.
}
In the one instanton case, the second term in 
equation \eqref{potential-F_01} was the only
place where the axion corresponding to the instanton appeared.
Recalling $T_i = \hat{\tau}_i + i \rho_i$, such a  term could be
written as $Xe^{i\rho}+\ov{X}e^{-i\rho}$ and upon
minimising the potential with respect to  $\rho$, it was rendered real and
negative \cite{Balasubramanian:2005zx,Conlon:2005ki}. The negativity of this term was crucial for the existence
of the minimum of the F-term potential at exponentially large volume.

In the general case of more than one instanton, the first
 term in \eqref{potential-F_01}
also depends on the axions, provided the volume of the intersection locus of the
respective instanton-cycles is non-vanishing. In this case, a 
more careful analysis of $V_F$ 
is needed, which we leave for future work \cite{future}. 
Requiring though that 
\bea
   \mbox{Vol}\bigl( D_{\alpha} \cap D_{\beta} \bigr)=0
\eea
for all pairs of instantons with $\alpha\ne \beta$ , guarantees that the respective axions
are stabilised in the way described above by the second
term in \eqref{potential-F_01}.

For the following, we will restrict ourselves
to the case of one instanton wrapping a general four-cycle $D_{\rm E3}$ in the Calabi-Yau manifold.
Employing then the stabilisation of the axion associated to the instanton illustrated above, the F-term potential for one E3-brane 
instanton simplifies to
\eq{
  \label{potential-F_02}
  V_F \simeq &
    -\frac{(2\pi)^2}{\hat\V}\:
    \mbox{Vol}\bigl( D_{\rm{E3}} \cap D_{\rm{E3}} \bigr)\:
    \bigl| A_{\rm{E3}} \bigr|^2 \: e^{-4\pi\hat{\tau}_{\rm E3}}  \\
   &\hspace{50pt}-\frac{4\pi}{\hat\V^2}\:
     \hat{\tau}_{\rm E3}\: e^{-2\pi \hat{\tau}_{\rm E3}}\:
     \bigl| A_{\rm{E3}} W_0 \bigr|
    +\frac{3}{4} \:\frac{\hat\xi}{\hat\V^3}\: \bigl| W_0 \bigr|^2 .
}
This expression 
is nearly similar to the well-known
expression of $V_F$ (\ref{scapot}) in the original large volume scenario.
The only difference is the first term. If we find that
\bea
  \label{cond_intersection}
  \mbox{Vol}\bigl( D_{\rm E3}\cap D_{\rm E3} \bigr) \simeq 
  -\sqrt{\hat\tau_{\rm E3}} \;,
\eea
then as shown in \cite{Balasubramanian:2005zx}, we are  guaranteed to
find a minimum of $V_F$ at exponentially large values of $\V$ and with
$\tau_{\rm E3}\simeq\log (\V)$. However, in general the minima of $V_F$ will depend on the concrete model and on the way the moduli are stabilized.

Let us summarize the results of this part. Performing the large volume expansion
 of the scalar
F-term potential for a general instanton configuration leads 
to an expression where the axions corresponding to the instantons cannot be stabilized easily. 
We did not attempt to address this question but restricted us to the case of one instanton along
a general four-cycle.

\bigskip
The main question is now 
whether it is indeed possible to freeze the
K\"ahler moduli controlling the size of the MSSM 
D7-branes via the D-terms of the $U(1)$ gauge factors
supported on these D7-branes and whether these sizes
are of the same order of magnitude as the instantonic
four-cycles.
Let us collect the formal constraints we have to
successfully implement in a concrete model for this scenario to work:
\begin{itemize}
\item
   Find a Calabi-Yau of swiss-cheese type with
   one large four-cycle controlling the size of the manifold
   and small cycles typically arising from resolutions
   of singularities.\footnote{It would be interesting to investigate
   whether also for instance Calabi-Yaus with a fibration
   structure can lead to large volume moduli freezing.
   For these the volume can usually be brought to  the schematic form
   ${\cal V}=\tau_1\sqrt{\tau_2}-\sum_I \tau_I^{3\over 2}$.}
\item
  Define an orientifold projection of this space 
  leading to
  O7- and O3-planes and freeze the complex structure and
  dilaton moduli by $G_3$-form flux. This latter will
   contribute to the D3-brane tadpole.
\item
  Introduce a set of intersecting (magnetised) D7-branes
   supporting the chiral MSSM spectrum and a hidden D7-brane sector
   such that the D7- and D3-brane tadpole cancellation conditions
   are satisfied. Moreover, the D7-branes must be free of
   Freed-Witten anomalies \cite{Freed:1999vc}. 
\item
  Classify all E3-instantons on this space which from the  
   zero mode structure can contribute to the uncharged superpotential.
   For this, a sufficient condition is that the instanton is rigid
   and has no other chiral or vector-like  zero modes from E3-D7 
   intersections. 
   Furthermore, one also needs to ensure that 
   the instantons are free of Freed-Witten anomalies 
   \cite{KashaniPoor:2005si}.
\item
  Compute the effective F- and D-term potential and analyse whether
  the combination of both freezes all K\"ahler moduli inside the
  K\"ahler cone with the size of the D7-branes coming out of  the same 
  order
  as the sizes of the instantons $\tau\simeq \log({\cal V})$.
\end{itemize}
Moreover, since in the non-supersymmetric 
large volume minimum the D-terms vanish, we still only have
F-term supersymmetry breaking and the soft-terms
can be computed in the usual way \cite{Brignole:1997dp}.

In the remainder of this paper, we will explicitly
carry out some of the steps mentioned above for 
a concrete Calabi-Yau orientifold  model.
Our simple (toy) model is neither realistic nor can all
conditions mentioned above be met explicitly, but
it  nevertheless shows how 
this program can partly be realised 
even on a simple Calabi-Yau manifold. We leave a more phenomenological
discussion of this moduli freezing scenario for future work.


\section{The $\IP_{[1,3,3,3,5]}[15]$ Calabi-Yau}
\label{sec_example}

For the large volume scenarios reviewed in section \ref{sec_lvs} it is now clear 
that we need at least three K\"ahler moduli to have both E3-instantons and a 
chiral D7-brane sector. 
The exponentially large cycle, controlling the overall 
size of the manifold, is usually frozen by the competing effects of the 
leading order $\alpha'$-corrections to the K\"ahler potential and the
 E3-instanton contribution. On the small cycles of a swiss-cheese type 
Calabi-Yau, the instantons and the D7-branes will be distributed.

Checking some  Calabi-Yau three-folds defined as
hypersurfaces in weighted projective spaces, we found one candidate
which actually is of swiss-cheese type.
It is the resolution of the $\IP_{[1,3,3,3,5]}[15]$ manifold.
It will turn out that this Calabi-Yau is still not rich enough
to allow for complex structure moduli stabilisation by 
fluxes and a complete MSSM sector, but serves as
a simply toy model to give a proof of principle  how the combination
of F- and D-term moduli stabilisation can work in more
realistic models. 
Let us describe the algebraic geometry of this Calabi-Yau 
in some more detail in  the next subsections.


\subsection{The topology of $\IP_{[1,3,3,3,5]}[15]$}


\vspace{2mm}
\subsubsection*{Toric resolution}

The $\IP_{[1,3,3,3,5]}[15]$ manifold has a $\mathbb Z_3$ singularity 
along the complex line $x_1=x_5=0$, which is met by the
hypersurface constraint. The  resolution of this $A_2$ orbifold singularity 
introduces two intersecting $\IP^1$s over the line. 

This resolution is easily described invoking the methods of toric geometry.
Besides the five divisors $v^*_1=(1,0,0,0)$, $v^*_2=(0,1,0,0)$, 
$v^*_3=(0,0,1,0)$,  $v^*_4=(0,0,0,1)$, $v^*_5=(-3,-3,-3,-5)$  
one introduces the two blowing-up
divisors  $v^*_6=(-2,-2,-2,3)$ and  $v^*_7=(-1,-1,-1,-1)$. 
The unique maximal triangulation is then given by
\eq{
  {\rm Triangle}=& \bigl\{ [1,2,3,4], 
  [1,2,3,5],[1,2,4,7],[1,2,6,7],[1,2,5,6],
  [1,3,4,7],\\
  &\ \  [1,3,6,7],[1,3,5,6],[2,3,4,7],[2,3,6,7],[2,3,5,6] \bigr\}
  \;.
}
The data
of the associated linear sigma model is the following. We have seven complex coordinates 
$x_i$  with three $U(1)$ symmetries. The corresponding charges are shown in \eqref{sigmam}.
\eq{
  \label{sigmam}
  \renewcommand{\arraystretch}{1.1}
  \renewcommand{\arraycolsep}{7pt}
  \begin{array}{|c|c|c|c|c|c|c||c|}
  \hline
  x_1 & x_2 & x_3 & x_4 & x_5 & x_6 & x_7 & p \\ \hline\hline
  3 & 3 & 3 & 5 & 1 & 0 & 0 & 15 \\
  2 & 2 & 2 & 3 & 0 & 1 & 0 & 10 \\
  1 & 1 & 1 & 1 & 0 & 0 & 1 & 5 \\
  \hline
  \end{array}
}
The divisors $D_i$ are defined by the constraints $x_i=0$ and 
the resulting Stanley-Reisner ideal reads \footnote{We used
the maple package ``Schubert'' to perform part of these computations.} 
\bea
  SR=\bigl\{\: x_4\, x_5\: ,\: x_4\, x_6\: ,\: x_5\, x_7\: ,\:
  x_1\, x_2\, x_3\, x_6\: ,\:  x_1\, x_2\,x_3\, x_7\: \bigr\} \;.
\eea
The triple intersection numbers in the basis  
$\eta_1=D_5$, $\eta_2=D_6$, $\eta_3=D_7$ are calculated as 
\bea
  I_3=9\,\eta_1^3 -40\, \eta_2^3 -40\,\eta_3^3-15\,\eta_1^2 \eta_2
  +25\,\eta_1 \eta_2^2 -5\,\eta_2^2 \eta_3 +15\,\eta_2 \eta_3^2
  \;. 
\eea
From section \ref{sec_lvs} we recall that the volume $\tau_i$ of the divisor 
$D_i$ and the overall volume of the manifold (in string-frame) are expressed in terms of the 
K\"ahler form in the following way
\eq{
  \label{ex_vol_four}
   \tau_i 
  = \frac{1}{2}\int_{\cal X} \bigl[ D_i \bigr]\wedge J \wedge J \;,
  \hspace{30pt}
  \V = \frac{1}{6} \int_{\cal X} J\wedge J\wedge J\;.
}
Expanding then the K\"ahler form in the basis $\{\eta_1,\eta_2,\eta_3\}$ from 
above as $J=\sum_{i=1}^3 t_i\, [\eta_i]$ we find for the volumes of the divisors 
$D_5$, $D_6$ and $D_7$
\eq{
  \tau_5 &= \frac{1}{2}\: \Bigl( 3\,t_1-5\,t_2 \Bigr)^2 \;,\\
  \tau_6 &= \frac{5}{6}\: \left[ \Bigl( 3\,t_3 -t_2 \Bigr)^2-
  \Bigl( 5\,t_2 -  3\,t_1 \Bigr)^2  \right]   \;,\\
  \tau_7 &= -\frac{5}{2}\: \Bigl( t_2-4\, t_3 \Bigr) \, 
      \Bigl( t_2-2\, t_3 \Bigr) \;.
}


\subsubsection*{The K\"ahler cone}

Next, we are going to determine the K\"ahler cone, which
is defined by the condition that the volumes of all effective
curves $\mathcal{C}$ are positive.
The first step is to compute the cone of all effective curves, which
is called the Mori cone and then deduce from this the K\"ahler cone
by the condition  $\int_{\mathcal{C}} J >0$. 
The resulting constraints describing the K\"ahler cone are 
\eq{
  t_2-2\, t_3 >0 \;,\qquad
  t_1-2\, t_2+t_3 >0 \;,\qquad
  -3\, t_1+5\, t_2>0 \;.
}
These conditions ensure also that the overall volume $\V$ is positive, 
that all volumes of effective divisors are positive and, by construction, 
that all volumes of holomorphic curves are positive.


\subsubsection*{Swiss-cheese structure}

For a large volume compactification we want to make one four-cycle large 
while keeping the others small. Let us therefore take a closer look at the 
volume. Using the K\"ahler cone restrictions above, we find that $\V$ can
 be written as
\eq{
  \label{vol1}
  \V = \sqrt{\frac{2}{45}} \left(
  \bigl( 5\tau_5 + 3\tau_6 + \tau_7 \bigr)^{3/2}
  - {\textstyle \frac{1}{3}}\, \bigl( 5\tau_5 + 3\tau_6 \bigr)^{3/2}
  - {\textstyle \frac{\sqrt{5}}{3}}\, \bigl( \tau_5 \bigr)^{3/2}
  \right) \;.
}
From this expression we see that this model admits a swiss-cheese structure. 
Indeed, we can make $\tau_7$ large so that the total volume $\V$ becomes 
large while keeping the four-cycles volumes $\tau_5$ and $\tau_6$  small. On the latter ones the D-branes supporting the MSSM will be wrapped.

In such a setup, we are thus not allowed to wrap D-branes supporting
the MSSM on (some combination involving) 
$D_7$ because then the gauge coupling $1/g_{\rm YM}^2\sim \tau_7$ would be 
too small. Similarly, we ignore instantons along this divisor, 
because  its contribution to the superpotential is
exponentially suppressed.
We are then left with the two divisors $D_5$ and $D_6$. 
Note however that not all combinations of $D_5$ and $D_6$ are allowed. 
We have to wrap D-branes and instantons along effective cycles,
 i.e. positive linear combinations of the divisors.


\subsubsection*{Rigid cycles}

Furthermore, we require  the instanton  to be rigid in the sense  that
no extra fermionic zero modes from the deformations
of the cycle or from Wilson lines along one-cycles do appear. 
The transverse deformations of a holomorphic four-cycle $D$ 
are counted by the global sections of the normal bundle $N$ of $D$.
By the adjunction formula and Serre duality on $D$ we get
$H^0(D,N_D)=H^2(D,{\cal O}_D)$.
The Wilson lines are counted by the non-contractable one-cycles
on $D$, which are counted by $H^1(D,{\cal O}_D)$.
Therefore, for an instanton to not  have additional deformation
zero modes we will require
\bea
  H^0\bigl(D,{\cal O}_D\bigr)=1\;,
  \qquad H^{i}\bigl(D,{\cal O}_D\bigr)=0 \;,
  \qquad {\rm for}\quad i=1,2\; .
\eea
A necessary  criterion for this is that the Euler characteristic
of the trivial line bundle over $D$ is equal to one, i.e.
\bea
   \chi\bigl(D,{\cal O}_D\bigr)
   =\sum_{i=0}^2 (-1)^i\, H^i (D,{\cal O}_D)=1 \;.
\eea
Employing the Koszul sequence
\bea
\label{koszul}
 0 \rightarrow {\cal O}_{\cal X}[-D] \rightarrow {\cal O}_{\cal X}
 \rightarrow {\cal O}_D \rightarrow 0 \;,
\eea
and the resulting long exact sequence in cohomology,
one obtains  the relation $\chi(D,{\cal O}_D)=\chi({\cal X},{\cal O}[-D])$.

In our concrete example, for a four cycle $D=m\, \eta_1+n\, \eta_2+l\, \eta_3$ the Euler characteristic is calculated as
\eq{
  \label{euler}
  \chi\bigl(D,{\cal O}_D\bigr)
  = {\textstyle \frac{15}{2}} nl^2 + {\textstyle\frac{25}{2}} mn^2 
  - {\textstyle\frac{5}{2}} n^2l&-{\textstyle\frac{15}{2}} m^2n
  -{\textstyle\frac{20}{3}}l^3 \\
  &- {\textstyle\frac{20}{3}} n^3  
  + {\textstyle\frac{5}{3}}n + {\textstyle\frac{5}{3}} l + 
  {\textstyle\frac{3}{2}}m^3-{\textstyle\frac{1}{2}} m \;.
}
Looking via a computer search for combinations with 
$\chi(D,{\cal O}_D)=1$ and $l=0$ we have found the solutions
\bea
\label{divs}
  (m,n,l) = \bigl\{ (1,0,0)\;,\;\;
            (1,1,0)\;,\;\;
            (2,1,0)\;,\;\;
            (2,2,0)\;,\;\;
            (12,11,0) \bigr\} \;.
\eea
In order to compute the precise cohomology classes $H^i(D,{\cal O}_D)$,
we use the cohomology classes of general line bundles on the toric ambient 
space shown in Appendix \ref{app_cohom}
and then run them through the Koszul sequences for the restrictions
on the Calabi-Yau hypersurface and the divisors $D$.
The result is that the first four divisors in (\ref{divs})
really have $H^i(D,{\cal O}_D)=(1,0,0)$, i.e. these are
irreducible effective divisors without any Wilson lines or transverse
deformations.

\begin{figure}[ht] 
\begin{center} 
 \subfigure[$D_5+D_6$]{
 \includegraphics[height=3.8cm]{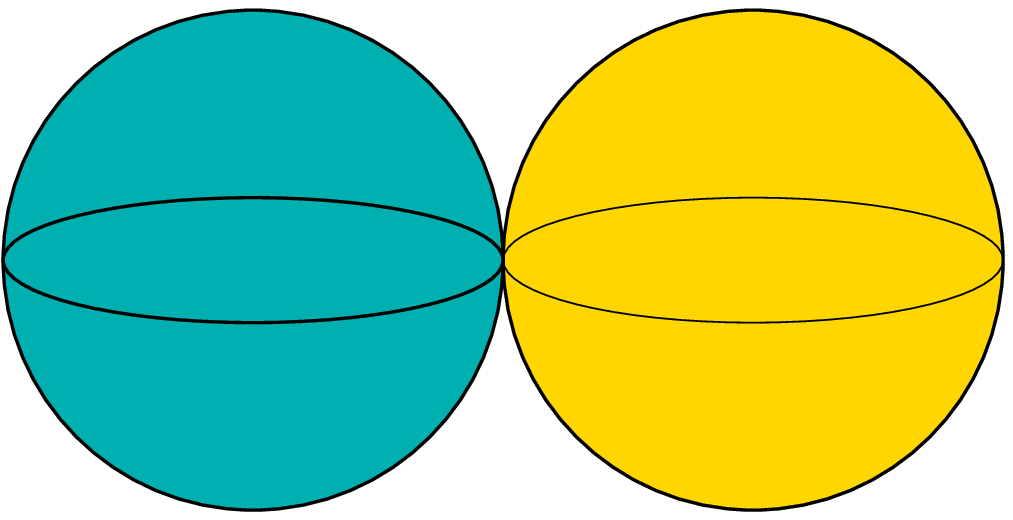} 
 \begin{picture}(0,0)
 \put(-189,8){{\large $D_5$}}
 \put(-21,8){{\large $D_6$}}
 \end{picture}
 \label{figspheres}
 }
 \hspace{1.5cm}
 \subfigure[$2D_5+D_6$]{
 \includegraphics[height=4.5cm]{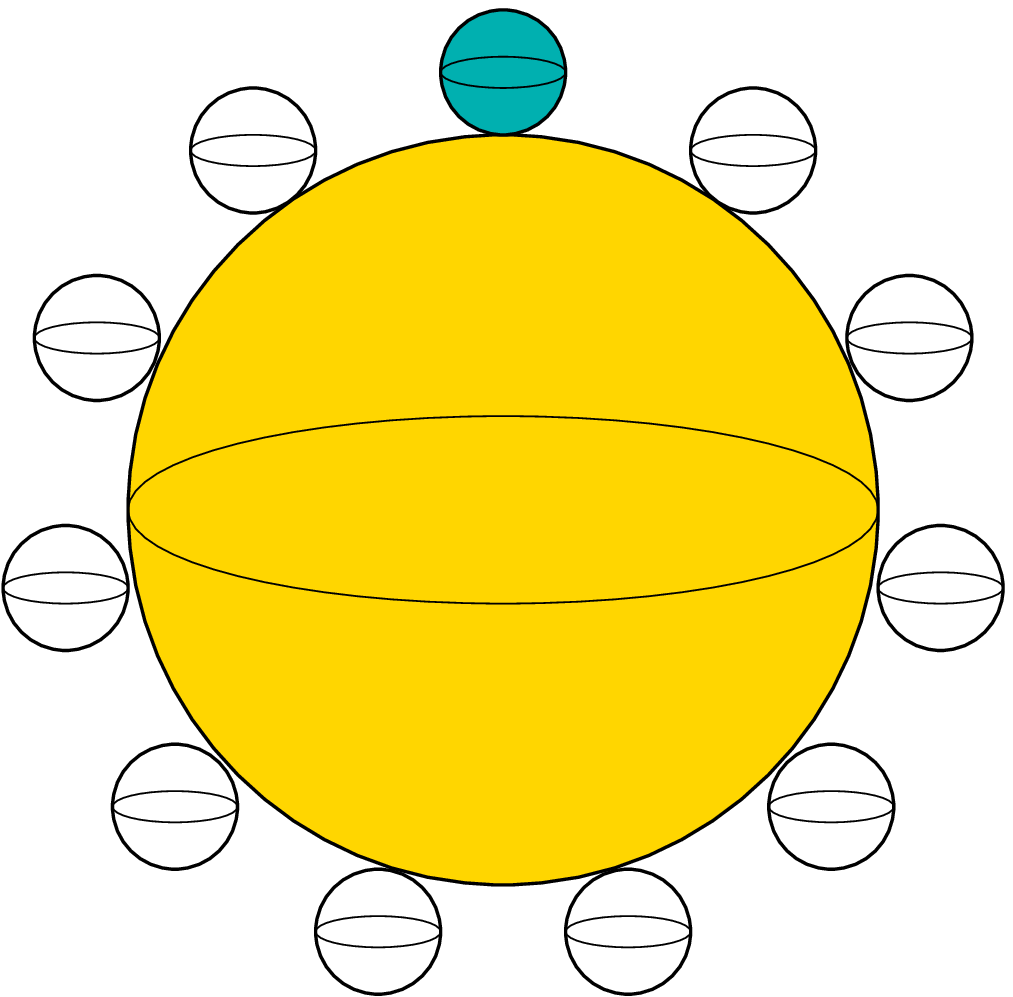} 
 \begin{picture}(0,0)
 \put(-55,125){{\large $D_5$}}
 \put(-80,30){{\large $D_6$}}
 \end{picture}
 \label{figinst} 
 }
\end{center} 
\vspace{-5mm}
\caption{\small Singular rigid divisors.} 
\end{figure}

\bigskip
One comment is in order here. Note that the three rigid divisors
$(1,1,0)$, $(2,1,0)$, $(2,2,0)$ are singular. Let us explain this for
the first one $D_5+D_6$. The only constraint one can write down of this degree
is $Q=x_5\, x_6 =0$. This defines two complex divisors $x_5=0$ and
$x_6=0$ intersecting along the  curve $x_5=x_6=0$, where
the manifold becomes singular. Since the four-cycle has no deformations,
the singularity cannot be smoothed out. A lower dimensional 
analogy is shown in figure \ref{figspheres}.
In the following we allow E3-instantons and D7-branes to also
wrap these rigid cycles \footnote{We have been informed by Volker
Braun, that they have identified  such topologies of world-sheet instantons
to contribute to the heterotic superpotential \cite{Braun:2007tp,Braun:2007xh,Braun:2007vy}.}.
Once we will have specified our orientifold projection, we will show
that all these rigid cycles carry $SP$ gauge group for D7-branes
wrapped around them and consequently $SO$ gauge group
for wrapped E3-instantons.

A similar structure appears for the rigid cycle
$2D_5+D_6$. By computing $H^i(2D_5,{\cal O}))=(11,0,0)$, we find that 
the instanton divisor has the structure shown figure \ref{figinst}.
The intersection between $D_6$ and any of the eleven components of 
$2D_5$ is over a two cycle. 

\subsubsection*{Diagonal basis}

In the following it will be more convenient to work in a basis where the 
volume $\V$ as well as the triple intersection numbers become particularly 
simple. Guided by \eqref{vol1},  we introduce the new basis of divisors as
\eq{
  D_a = 5\, D_5 + 3 \,D_6 + D_7\;, \qquad
  D_b = 5\, D_5 + 3 \, D_6 \;, \qquad
  D_c = D_5 \;, 
}
for which the triple intersection numbers diagonalise
\bea
   I_3= 5\, D_a^3 + 45\, D_b^3 + 9\, D_c^3\;.
\eea
The total volume in terms of the divisor volumes $\tau_a$, $\tau_b$ and $\tau_c$ reads
\eq{
  \V = \sqrt{ \frac{2}{45}}\left(
  \tau_a ^{3/2} - {\textstyle \frac{1}{3}} \tau_b^{3/2} 
  - {\textstyle \frac{\sqrt{5}}{3} } \tau_c^{3/2} \right) \;.
}
Expanding also the K\"ahler form in this diagonal basis as 
$J=t_a\, D_a-t_b \, D_b-t_c \, D_c$, we find that the K\"ahler cone 
conditions have the  very simple form
\eq{
  \frac{1}{3}\; t_a > t_b > t_c >0 \;.
}
As one can see from the  above, the large divisor is 
now simply $D_a$. For the gauge couplings not to be unrealistically small,
we do not wrap the D7-branes supporting the MSSM along the large cycle. Moreover, 
significant E3-instanton contributions only arise
from instantons wrapped on the small four-cycles.
Therefore, we can make the general ansatz for the 
D-brane and instanton cycles
\eq{
  \label{ansatz_cycles}
  D_{\rm{D7}} = n_b\, D_b + n_c\, D_c \;, \hspace{30pt}
  D_{\rm{E3}} = m_b\, D_b + m_c\, D_c \;, 
}
where now the wrapping numbers $n$ and $m$ need not be integer.
They are related to the wrapping numbers $n_i$  in the
$\{\eta_i\}$ basis  by 
\eq{
  n_b = \frac{1}{3}\, n_2\;, \hspace{30pt}
  n_c = n_1 - \frac{5}{3}\, n_2 \;,
}
and similarly for $(m_b,m_c)$.


\subsection{Moduli stabilisation}

Now that we have collected all the topological data, we can develop our model further. 
It will turn out that the tadpole cancellation conditions for the present setup impose strong restrictions so that  we 
cannot consider a full MSSM set-up but only a chiral toy model.
We will have two stacks of D7-branes wrapping rigid four-cycles $D_A$ and $D_B$ where only on the first one a non-trivial line bundle
${\cal L}_A$ is turned on. 
We consider the Standard Model as being part of the
$U(N_A)$ gauge group on the first stack of branes (even though
in the eventual model it will not have large enough gauge group). 
Then we get MSSM matter from the intersections $AA'$ and $AB$
where the prime denotes the orientifold image.
Connecting to our discussion in section \ref{sec_moduli_stab}, 
we in general allow the gauge group $U(N_A)$ to be larger
than just the MSSM gauge group. Then from the two intersections
$AA'$ and $AB$ we get matter $\Phi_{\rm SM}$ which is part
of the Standard Model. Furthermore, we get other matter $\Phi_{\rm abel}$
transforming in singlet representations of the MSSM gauge group, but
 carrying  certain charges under abelian $U(1)$s orthogonal
to $U(1)_Y$.
In addition, in order for satisfying the D7-brane tadpoles
we need extra hidden sector branes.
 
Before  we give the complete model, let us first elaborate
on the D- and F-term constraints. 


\subsubsection*{D-Term constraints}

In section \ref{sec_chiral_d7} we have explained that the D-terms in large 
volume scenarios should vanish in order not to spoil the $1/\V$ expansion 
of the scalar F-term potential and the resulting minimum. The D-term contains 
the Fayet-Iliopoulos parameter $\xi$ and the possible matter fields 
$\Phi_{\rm SM}$, $\Phi_{\rm exo}$ and $\Phi_{\rm abel}$. However, as 
argued previously, for the simple reason that the SM gauge symmetry
is unbroken at low energies, at least the VEVs of the first two 
matter fields have to vanish and for $\Phi_{\rm abel}$ it
is likely to vanish.
For the MSSM sector, we are thus left with the requirement that
$\xi_A=\xi_B=0$.
Recalling the precise form of the FI-parameter \eqref{FI_parameter}, the condition  $\xi_A=0$ reads
\eq{
  0 = \int c_1({\cal L}_A)\wedge 
  \bigl[ D_{\rm{D7}_A} \bigr]\wedge J
  \;.
}
For the second D7-brane the condition $\xi_B=0$ is trivially satisfied because of $c_1(\mathcal{L}_B)=0$. Next, we consider the (chiral) zero mode constraint from the D7-E3 intersections.  
The only non-trivial equation comes from D7$_A$ and reads
\eq{
  0 = \int c_1({\cal L}_A)\wedge 
  \bigl[ D_{\rm{D7}_A} \bigr]\wedge \bigl[ D_{\rm E3} \bigr] \;.
}
Using then our ansatz \eqref{ansatz_cycles} in the diagonal basis, we find that the only suitable solution to the two equations above is
\bea
       J=t_a\, \bigl[ D_a\bigr]  - t\,\bigl[ D_{\rm E3} \bigr]\;.
\eea

Let us note that this solution implies $t_b=\frac{1}{3}\,m_b\,t$ and $t_c=\frac{1}{3}\,m_c\,t$.
Comparing with the K\"ahler cone constraint $t_b>t_c>0$ and going back to 
the basis in $\{\eta_1,\eta_2,\eta_3\}$, we see that 
only wrapping numbers with $2m_2>m_1>\frac{5}{3}m_2$ are possible. 
This cannot be solved by any of the rigid cycles
$(m_1,m_2,m_3)\in \{ (1,0,0),(1,1,0), (2,1,0), (2,2,0)\}$. 
However, the  choice $(m_1,m_2,m_3)=(2,1,0)$, i.e. $D_{\rm E3}={1\over 3}(D_b+D_c)$,
is at least on the boundary of the 
K\"ahler cone at $t_b=t_c$. Of course, we cannot choose instantons at will but have to take all of them into account. But we can arrange our setup in such a way that only an instanton along the cycle $(m_1,m_2,m_3)=(2,1,0)$ contributes to the
stabilisation of the K\"ahler moduli. We will come back to this point after we specified the orientifold projection and the D-branes in our model.
Note furthermore, by allowing a non-vanishing VEV for
$\Phi_{\rm abel}$, it might be possible to fix $t_b$ and $t_c$ on a ray
inside the K\"ahler cone via the instanton above.

Let us now choose the stacks of D7-branes to wrap the rigid four-cycles
\eq{
  D_{\rm{D7}_A} = D_5 +  D_6 ={1\over 3}\:(D_b-2\, D_c)\;,\qquad 
  D_{\rm{D7}_B} = D_5 =D_c \;,
}
with the line bundles
\bea
   {\cal L}_A= {1\over 3}(2\, D_b  + 5\, D_c)\;,\qquad 
  {\cal L}_B={\cal O}  \;.
\eea
With this choice, as shown above, there are no chiral zero modes on the 
$D7-E3$ intersections. However, similar to \cite{Florea:2006si},
we expect both vector-like  
bosonic and fermionic zero modes, because, as shown in figure \ref{figinst}, 
the rigid E3-instanton actually contains both  $D_{\rm{D7}_A}=D_5+D_6$ and  
$D_{\rm{D7}_B}=D_5$ as a sub-locus.
One way to get rid of these zero modes, would be to turn
on discrete Wilson lines or discrete displacement on the D7-brane resp.
E3-instanton. It is beyond the scope of this paper to analyse
mathematically this possibility for these divisors.
From now on, we proceed by assuming that such non-chiral zero modes
can be made massive so that indeed the E3-instanton
on $D_{\rm E3}=2D_5+D_6$ contributes to the uncharged superpotential.

Before concluding this part, let us note that the vanishing of the D-term 
gives rise to a minimum of the scalar D-term potential. 
Moreover, we have argued that the F-term potential does not depend on at least 
one linear combination of K\"ahler moduli $\overline{\tau}$ which however 
appears in the D-term. For moduli stabilisation this means that
$\partial V/\partial\ov{\tau}=\partial V_D/\partial\ov{\tau}=0$ is solved by 
the vanishing D-term and thus in our setup fixes 
\eq{
  \label{kaehler_fix1}
  t_b = t_c =: t\;. 
}
In the diagonal basis this solution implies that $D_6$ shrinks to zero size but $D_5$ stays finite. Note first, our Standard Model branes do both involve $D_5$ and so their volume is always non-zero. Second, for a non-vanishing VEV of $\Phi_{\rm abel}$ we expect the volume of $D_6$ to be finite.


\subsubsection*{F-Term constraints}

Let us now go on and study the F-term potential. Since we only have 
a single
instanton contributing to the potential, we can refer to equation \eqref{potential-F_02}. Using then the concrete
data of our model, we find $\mbox{Vol}\bigl(D_{\rm E3} \cap D_{\rm E3} \bigr) = - 5t_b -  t_c$ and
therefore 
\eq{
  \label{numerics_1}
  V_F \simeq 
    \frac{(2\pi)^2}{\hat\V}\:
    \bigl( 5 \hat t_b + \hat t_c \bigr)
    \:
    \bigl| A_{\rm{E3}} \bigr|^2 \: e^{-4\pi\hat{\tau}_{\rm E3}}  
   -\frac{4\pi}{\hat\V^2}\:
     \hat{\tau}_{\rm E3}\: e^{-2\pi \hat{\tau}_{\rm E3}}\:
     \bigl| A_{\rm{E3}} W_0 \bigr|
    +\frac{3}{4} \:\frac{\hat\xi}{\hat\V^3}\: \bigl| W_0 \bigr|^2 .
}
The first term cannot be expressed as a square root of $\hat\tau_{\rm E3} = \frac{1}{6} \bigl( 45 \hat t_b^2 + 9 \hat t_c^2 \bigr)$ and so the analysis
of \cite{Balasubramanian:2005zx} for the minimum of $V_F$ at large volumes is
not applicable.
However, employing equation \eqref{kaehler_fix1},
we find the following relation between the volume of the instanton cycle and the volume 
of its self-intersection
\eq{
   \mbox{Vol}\bigl(D_{\rm E3} \cap D_{\rm E3} \bigr) = -6\:\hat t = -2\: 
   \sqrt{\hat\tau_{\rm E3}} \;.
}
Note that this volume formally is negative, which simply reflects
the fact that the four-cycle $D_{\rm E3}$ is exceptional with 
a  self-intersection not corresponding to an effective two-cycle.  
Using this relation, the above expression becomes
\eq{
  V_F \simeq 
    \frac{8\pi^2}{\hat\V}\:
    \sqrt{ \hat\tau_{\rm E3}}
    \:
    \bigl| A_{\rm{E3}} \bigr|^2 \: e^{-4\pi\hat{\tau}_{\rm E3}}  
   -\frac{4\pi}{\hat\V^2}\:
     \hat{\tau}_{\rm E3}\: e^{-2\pi\hat{\tau}_{\rm E3}}\:
     \bigl| A_{\rm{E3}} W_0 \bigr|
    +\frac{3}{4} \:\frac{\hat\xi}{\hat\V^3}\: \bigl| W_0 \bigr|^2 .
}
Recalling our discussion in section \ref{sec_scapot_fterm}, the $1/\hat\V$ expansion of the F-term potential is of the form which allows for a minimum of  $V_F$ at large values of $\hat\V$.

We can then treat these variables as fixed and use their relation to the 
K\"ahler moduli. We obtain 
\eq{
  t_b=t_c=t =  \frac{1}{3} \sqrt{ \tau_{\rm E3}}\;,\qquad
  t_a = \biggl(\: \frac{6}{5}\:\V_0 + \frac{2}{5}\: \tau_{\rm E3}^{3/2} 
  \:\biggr)^{1/3} \;,
}  
where we denoted the value of $\V$ in the minimum by $\V_0$.
Therefore, in this model all K\"ahler moduli have been stabilised. 
To be more precise, we have seen that all coefficients $t_a$ in the expansion 
of $J$ are fixed and so are the real parts of the K\"ahler moduli $T_i$. Furthermore, 
through the F-term potential the axion corresponding to the instanton cycle is 
stabilised and via the D-term and Green-Schwarz mechanism the axion 
associated with the matter sector gets massive. 

For the K\"ahler moduli, we now get three different mass scales.
Since the D-term vanishes in the minimum, the mass of the large volume
modulus and the small cycle fixed by the instanton do not change. 
Just keeping track of the $1/{\cal V}_0$ factor they scale like
$M_{\tau_b}\simeq M_{\rm Pl}/{\cal V}_0^{3/ 2}$ and
$M_{\tau_s}\simeq M_{\rm Pl}/{\cal V}_0$ \cite{Conlon:2005ki}.
The orthogonal K\"ahler modulus fixed by the D-term then has mass
$M_{\tau_D}\simeq M_{\rm Pl}/\sqrt{{\cal V}_0}$, which being of string scale size
is much heavier than
the other two.


\subsubsection*{Numerical analysis}

In order to explicitly check that the large volume
minimum of the full scalar potential persists in our model,
we have numerically evaluated equation \eqref{numerics_1}. 
Installing the appropriate factors of $2\pi$ and $g_s$, and choosing 
$|A_{E3}|=1$, $|W_0|=5$, we
minimised the function \footnote{A very similar potential appeared in
\cite{Conlon:2006tq}, but without the D-term part.}
\eq{
  \label{numerical_V}
  V_{F+D}\bigl(\,{\cal V},\tau_b,\tau_c\,\bigr) 
  =&+ \frac{18.6}{{\cal V}}\: 
  \bigl(\,\sqrt{5\tau_b}+\sqrt{\tau_c}\,\bigr)\:g_s\:
  e^{-\frac{4\pi}{3}\frac{1}{g_s}(\tau_b+\tau_c)} \\
  &- \frac{20.9}{{\cal V}^2}\:\bigl(\,\tau_b + \tau_c\,\bigr)\:
  g_s^2\:
  e^{-\frac{2\pi}{3}\frac{1}{g_s}(\tau_b + \tau_c)}
   +\frac{6.5}{{\cal V}^3}\: g_s^3 \\
  &+\frac{13.3}{\V^2}\:\frac{1}{\tau_b - 2\tau_c}\:g_s^3\:
  \bigl(\, \sqrt{5\tau_c} - \sqrt{\tau_b}\, \bigl)^2 \;.
}
Note that we have not yet fixed the value of $g_s$ which is determined 
by the VEV of the dilaton. We have assumed that
it is stabilized by fluxes and since we did not perform
an explicit analysis of this mechanism, we choose $g_s=1/10$ for convenience.
However, as noted in \cite{Conlon:2006wz}, the stabilised volume $\V$ will depend
exponentially on $g_s$ through $\V\sim e^{c/g_s}$ where $c$ is some constant. Thus, a more careful analysis of the flux sector is 
inevitable.

\begin{figure}[p]
\begin{center} 
 \includegraphics[width=0.7\textwidth]{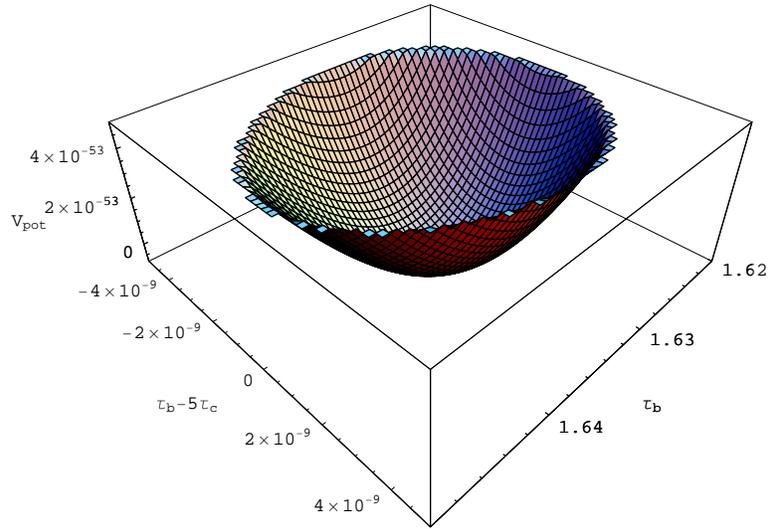} 
\end{center} 
\caption{{\small The potential $V({\cal V},\tau_b,\tau_{c})$ for $\V=2.15\cdot 10^{16}$.}\label{figpoti} } 
\end{figure}

\begin{figure}[p] 
\begin{center} 
 \includegraphics[width=0.7\textwidth]{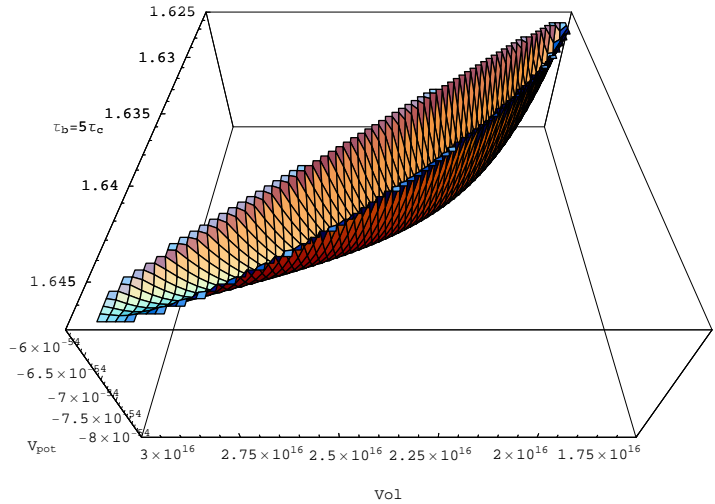} 
\end{center} 
\caption{{\small The potential $V( {\cal V}, \tau_b, \tau_c )$ for $\tau_c=0.33$.}\label{figpotii} } 
\end{figure}

Coming back to the potential above, we observe that the dominant part of 
\eqref{numerical_V} is given by the D-term potential fixing the combination 
$\tau_b=5\tau_c$. On top of that direction, we found a minimum of the 
potential in the variables $\V$ and $\tau_b$.
In figures \ref{figpoti} and \ref{figpotii}, we have plotted two sections 
through the parameter space showing the potential in the vicinity of the 
minimum.  The numerical values (in string units) in the minimum are 
${\cal V}\approx 2.2\cdot 10^{16}$ and the four-cycle volumes are stabilised 
at $\tau_b \approx 1.63$, $\tau_c \approx 0.33$.\footnote{If we minimise the potential \eqref{potential-F_00} instead of its large volume expansion \eqref{numerical_V}, we find the minimum at $\tau_c\simeq 0.53$, $\tau_b\simeq2.64$ and $\V\simeq 1.1\cdot 10^{13}$ for $g_s=1/10$. The difference in the value of $\V$ can be compensated
by arranging $g_s=1/12$ so that the minimum is at $\tau_c\simeq 0.53$, $\tau_b\simeq2.64$ and $\V\simeq 7\cdot 10^{15}$.}
  For the volume of the 
Standard Model cycles we find $\tau_{\rm SM}\simeq 0.33$ and the value
of the scalar potential in the minimum is of the order
 $V_{\rm min}\simeq -10^{-54}\, M_{\rm Pl}^4$. 

The stabilised four-cycle volumina are in a region where we
have to worry whether we can trust the supergravity
approximation.
Let us investigate more closely what the numerical reason is.
Recall from \cite{Balasubramanian:2005zx} the approximate formulas for the 
volume $\V$ and the four-cycle in the minimum 
\eq{
  \V \simeq 
  \frac{\mu\:g_s\:\bigl|W_0\bigr|}{2\:\lambda\:a_s\bigr|A_s\bigr|} 
  \left( \frac{4\:\nu\lambda\,\xi}{\mu^2}\right)^{1/3}
    e^{\frac{a_s}{g_s} \left( 
    \frac{4\:\nu\lambda\,\xi}{\mu^2}\right)^{2/3} }
    \;, \qquad 
    \tau \simeq \left( \frac{4\:\nu\lambda\,\xi}{\mu^2}\right)^{2/3} \;,
}
where we have used the notation from equation \eqref{scapot}. Note that 
$\lambda$ contains the information about the intersection of the instanton 
cycles and thus depends on the topology of the manifold and on the cycles 
suitable for instantons. Furthermore, $\xi$ is proportional to the Euler 
characteristic $\chi$ and so the above formulas depend
strongly on the topology of the compactification manifold.

For our present model, using the data after D-term fixing but leaving the 
Euler characteristic $\chi$ and the string coupling $g_s$ unspecified, 
we obtain
\eq{
\label{numeric}
  \V \simeq 6.1\cdot 10^{-2} \: g_s\: \bigl( - \chi \bigr)^{1/3}
  e^{0.145\: \frac{1}{g_s}\: (-\chi)^{2/3}} \;,
  \;\qquad
  \tau_{\rm SM} \simeq 1.2\cdot 10^{-2} \: \bigl( -\chi \bigr)^{2/3} \;.
}
Therefore the prefactor of order $10^{-2}$ in (\ref{numeric}) and the 
smallness of  the Euler characteristic $\chi=-144$ of our Calabi-Yau manifold
are the reasons for the string-frame four-cycle volume 
$\tau_{\rm SM}$ to come out so small.

Just as a rough estimate, let us analyse for which values of
$g_s$ and $\chi$ the formulas (\ref{numeric}) give more realistic
values of the K\"ahler moduli.
Choosing for instance $\tau_{\rm SM}=1.2$ leads to
$\chi\simeq -1000$. For the string coupling $g_s={3\over 8}$
we then get ${\cal V}=5\cdot 10^{15}$.
This points towards choosing Calabi-Yau's with Euler-characteristics
just at the limit of presently known examples for $\chi$.


\subsection{Orientifold with tadpole cancellation}

We now show that the setup introduced in the previous section can 
really be implemented in a globally defined orientifold model.
We choose the holomorphic involution 
$\sigma$ of the orientifold projection to permute the complex coordinates  $x_1$ and $x_2$. 
Then the divisors $D_5$, $D_6$, $D_7$ 
are invariant and we have $h_{1,1}^+=3$ and $h_{1,1}^-=0$. 
The orientifold plane is given by
\eq{
  D_{\rm{O7}} = 3 \eta_1 + 2\eta_2 + \eta_3 = D_a -\frac{1}{3}\, 
  D_b - \frac{1}{3}\, D_c \;,
}
and the cohomologies for this cycle are $H^i(D_{\rm{O7}},\mathcal{O})=(1,0,3)$. 
A careful analysis shows that in addition,  
the orientifold projection also leaves three points on the Calabi-Yau manifold
invariant. This can be seen as follows.
Choose the intersection of $x_3=x_4=x_7=0$ which gives five points.
These are described as the solutions to the equation $x_1^5+x_2^5=0$ 
in the variables $(x_1,x_2,x_5,x_6)$ up
to the projective identifications shown in \eqref{sigmam}.
These latter allow to fix $(x_5,x_6)$ to the point $(1,1)$ and
to see that the solution $(x_1,x_2)=(1,-1)$ is invariant under $\sigma$. 
The other four points are pairwise interchanged. 
The same story also holds on the intersections  $x_3=x_6=x_7=0$
and  $x_3=x_5=x_6=0$ giving the claimed three O3-planes.

Let us state a criterion by which we can decide whether on 
a stack of $\Omega\sigma (-1)^{F_L}$ invariant D7-branes
we get an $SO$ or $SP$ projection.
In the geometric orientifold we are considering, placing the D7-branes
right on top of the O7-plane gives an $SO(8)$ gauge symmetry. 
Wrapping another D7-brane on a $\sigma$ invariant four-cycle $D$ with trivial
gauge bundle, also leads to an $SO$ or $SP$ gauge symmetry.
If the configuration is such that $D$ intersects the O7-plane
over a two-cycle, then locally around the intersection the open string
stretched between the D7-brane on $D$ and the one on $D_{\rm O7}$
has four Neuman-Dirichlet boundary conditions and therefore carries
an $SP$ gauge group.

\medskip
Let us first consider  the simple model with eight D7-branes on top of the O7-plane.
Using $\int_{D_{\rm{O7}}}  \mbox{c}_2\bigl( T_{\rm{O7}} \bigr) = 45$
in the eq. (\ref{drauf}),  the resulting D3-brane tadpole is
\bea
    N_{\rm D3}+N_{\rm flux}=12 \;,
\eea 
so that the Euler characteristic of the Calabi-Yau four-fold
is $\chi({\cal Y})=288$.

Now we are considering the model from the previous section with 
magnetised D7-branes. 
For convenience, let us recall its instanton and D7-brane data 
\eq{
  \renewcommand{\arraystretch}{1.3}
  \begin{array}{l@{\hspace{3pt}}c@{\hspace{3pt}}l@{\hspace{3pt}}c@{\hspace{3pt}}lc
  l@{\hspace{3pt}}c@{\hspace{3pt}}l}
  D_{\rm{E3}} & = &2\eta_1 + \eta_2 &=& \frac{1}{3} \bigl( D_b + D_c 
  \bigr) \;,
    &\quad & c_1\bigl(\mathcal{L}_{\rm{E3}}\bigr) &=& 0\;, \\
  D_{A} & =& \eta_1+\eta_2  &=& \frac{1}{3} \bigl( D_b - 2 D_c \bigr) 
  \;,
    &\quad & c_1\bigl(\mathcal{L}_{A}\bigr) &=& 5\eta_1+2\eta_2 = 
    \frac{1}{3}\bigl( 2D_b+5D_c \bigr)\;, \\
  D_{B} & =& \eta_1  &=& D_c \;,
    &\quad & c_1\bigl(\mathcal{L}_{B}\bigr) &=& 0 \;.
  \end{array}
}
Note that since we do not have a gauge bundle on the second D7-brane, 
it is invariant  under the orientifold projection. 
Because it intersects the O7-plane over a two-cycle,  according to
our criterion from above it carries an $SP(2N_B)$  gauge symmetry.
Similarly, the $\Omega\sigma (-1)^{F_L}$ invariant
instanton cycle $D_{\rm E3}$ intersects the O7-plane over a
two-cycle and carries therefore an $O(1)$ gauge symmetry.

The chiral matter between the D7-branes can be computed using the rules from 
table \ref{tab_chir_spec}. Leaving the number of coincident branes $N_A$ and 
$N_B$ unspecified, we find
\eq{
  10 \times \bigl[ \overline{A}_A \bigr] + 10 \times \bigl[ 
  \overline{S}_A \bigr]
  + 10 \times \bigl[ (N_A,2N_B ) \bigr] \;.
}
However, this spectrum is only free of anomalies if we impose $N_B=N_A$. 
Thus, the spectrum of our model reads as follows.
\eq{
  \begin{array}{l@{}c@{}c@{}c@{}r}
  & U(N_A) &\;\times\;& Sp(2N_A) \\ \hline\hline \\[-4mm]
  10\times\bigl( &  \overline{ 
  \begin{array}{@{\hspace{1pt}}l@{\hspace{1pt}}}\\[-6mm]\Box\\[-2.27mm]\Box\end{array} } &,& 1&\bigr) \\
  10\times\bigl( &  \overline{\Box\hspace{-0.65mm}\Box} &,& 1&\bigr) \\
  10\times\bigl( &  \Box &,&\Box&\bigr)
  \end{array}
}
Furthermore, we have to satisfy the tadpole cancellation condition for the 
D7-branes which restricts $N_A$ as $N_A \leq 6$.

The D3-brane tadpole is more involved. The various topological quantities contributing 
to the formula \eqref{tapole_d3} are found as
\eq{
   \int_{D_A} \mbox{ch}_2\bigl( \mathcal{L}_A \bigr) = -5\;, \quad
   \int_{D_A}  \mbox{c}_2\bigl( T_{D_A} \bigr)  = -17\;, \quad  
     \int_{D_B}  \mbox{c}_2\bigl( T_{D_B} \bigr) =3\;. 
}
The condition is that, after including also the hidden sector branes,
the number of $N_{\rm D3}+ N_{\rm flux}$ is non-negative. 
We found only one solution, which works.
We choose the minimal case $N_A=1$ and in order to satisfy the D7-brane
tadpole constraint, we include the hidden branes 
\bea
       &&N_C=3:\ \    D_C=D_{\rm O7}=3\eta_1+2\eta_2+\eta_3\;,\\
      && N_D=1: \ \    D_D=\eta_1+\eta_2+\eta_3\nonumber\;,
\eea
with trivial gauge bundles.  The four-cycle $D_C$ is equal to the
O7-plane and $D_D$ is a rigid cycle with $H^i(D_D,{\cal O})=(1,0,0)$
and $\int_{D_D}  \mbox{c}_2\bigl( T_{D_D} \bigr) =-37$. 
Adding up all contributions to the D3-brane tadpole condition gives
\bea
  \label{tadpole_d3flux}
  N_{\rm D3}+ N_{\rm flux}=3\;.
\eea

After having specified the orientifold projection and the D-branes in our model, we can now revisit our claim that only the $O(1)$ instanton along $2D_5+D_6$ contributes to the stabilisation of the K\"ahler moduli. 
Using equation \eqref{intersection_iib}, we find that there are always
chiral zero modes between stringy instantons and D-branes except for $D_{\rm E3}=2D_5+D_6$. Therefore, if present, a term in the superpotential involving the Standard Model fields will be generated. Following our argumentation from section \ref{sec_moduli_stab}, such contributions have to be absent because Standard Model fields should not acquire a VEV. Similarly, the contribution from the gauge instanton on top of $D_A$ has to vanish. Thus, only the instanton $D_{\rm E3}=2D_5+D_6$ will contribute to the stabilisation of the K\"ahler 
moduli. However, we have to emphasize that actually also the vector-like instanton zero modes have to be determined as well as other mechanism to soak up unwanted fermionic zero modes have to be checked.

\bigskip
To conclude, it is clear that the constraint \eqref{tadpole_d3flux} might not give enough freedom for the three-form fluxes
to freeze all complex structure moduli.
For also including this sector consistently, one needs more
involved Calabi-Yau spaces.  
However, we have demonstrated  at a specific swiss-cheese Calabi-Yau manifold
with $h_{1,1}(\mathcal{X})=3$ that a combination of E3-instantons and D7-brane
D-terms can fix all three K\"ahler moduli in the large volume 
regime with all small cycles wrapped by D7-branes of order $\log({\cal V})$.
In our case, the D-terms (only) fixed the moduli on the boundary of the 
K\"ahler cone, where the four-cycle $D_6$ collapses. 
Furthermore, we have argued that out of the rigid divisors (\ref{divs}) only $D_{\rm E3}=2\eta_1+\eta_2$ contributes to the uncharged superpotential.
However, actually the complete 
vector-like zero mode spectrum has to be computed for such 
overlapping singular divisors and presumably also discrete
Wilson lines and displacements have to be included.
This complete mathematical investigation for this specific model 
is beyond the scope of this paper, whose punchline is rather
to exemplify for a concrete Calabi-Yau  that the F- and D-term freezing
scenario  has a good chance to be realisable in concrete large volume 
Type IIB orientifolds with a chiral D7-brane sector.


\section{Conclusions}

In this paper we have analysed the problem of combining K\"ahler 
moduli stabilisation by instantons resp. gaugino condensation with
a chiral D7-brane sector carrying the unbroken chiral
gauge theory which we would like to have in four dimensions.
Clearly, in order to make progress in deriving viable and
predictive string compactifications, this question is of
utmost importance. 

We argued quite generally, 
employing both string consistency conditions as well as phenomenological
input, that for chiral D7-brane sectors only a combination
of F- and D-terms can fix all K\"ahler moduli.
Then we investigated whether for the very promising large
volume scenario all their unquestionable nice features can 
be preserved once these D-terms are taken into account.
We showed that for more than one E3-instanton
also the F-term scalar potential contains new terms containing
the axionic fields, which potentially destabilise the large volume scenario.
Requiring these terms to be absent means that the instanton
cycles should not intersect.
Moreover, we also allowed for singular four-cycles, which homologically are 
linear combinations of the elementary ones.
These also induce a different moduli dependence in the 
F-term scalar potential.
We plan to investigate the general consequences of such many-instanton contributions in a future work \cite{future}.

In this paper we exemplified our general arguments about F-and D-terms
by  constructing a concrete
Type IIB orientifold on a (new) swiss-cheese type Calabi-Yau
manifold with three K\"ahler moduli. Ignoring the details of the
three-form flux sector, we constructed a global tadpole cancelling model
which showed all the features we do expect for a realistic 
model. We had a chiral intersecting D7-brane sector and a sector
of hidden branes filling up the D7-brane tadpole constraint.
Due to chirality there was an induced D-term, fixing (for vanishing
VEVs of matter fields) one combination of the K\"ahler moduli
at the boundary of the K\"ahler cone.
We had one rigid small cycle unoccupied by the D7-branes, so that
a stringy  $O(1)$ E3-instanton
wrapped on this cycle contributed to the superpotential.
Then the F- and D-terms together fixed the overall volume ${\cal V}$
at large values and the two diagonal small ones at 
size $\tau_i\simeq \log({\cal V})$ 
in a such a way that another four-cycle collapsed. 
Of course there will be world-sheet instanton corrections from this
collapsed cycle as well as probably also non-negligible string loop corrections, 
but we only expect them to contribute
to the K\"ahler potential (effectively changing $\xi$) and
to the Fayet-Iliopoulos terms, such that the four-cycle
volume is stabilised at order $\tau\simeq \ell_s^4$. 
Note that, if eventually  some of the $\Phi_{\rm abel}$ matter fields
are fixed at non-zero value, the D-terms can freeze
the K\"ahler moduli inside the K\"ahler cone.

We consider our simple toy model as a proof of principle that the LVS can
be robust enough that also chiral D7-brane sectors can be 
introduced. Of course, phenomenologically  our model
is not satisfying yet. The gauge group and matter content is 
not realistic and the D3-brane tadpole constraint leaves
probably not enough freedom to fix all complex structure
moduli by three-form fluxes. Moreover, our analysis of the
non-chiral zero modes was not complete.
However, we are confident that these shortcoming only
reflect the simplicity of the used Calabi-Yau space.
Using orientifolds of Calabi-Yau manifolds, for which 
the D3-brane tadpole is much larger than $\chi({\cal Y})=288$
will remedy  these problems. To this end, it would be very important
to know which of the toric Calabi-Yau manifolds  in the list of \cite{Kreuzer:2000xy}
have a swiss-cheese like structure, respectively can lead
to large volume moduli stabilisation.
It might be technically very challenging \footnote{{\it `\hspace{1pt}Quam multa fieri non posse, priusquam sint facta, iudicantur?\hspace{1pt}'} [Plinius, Naturalis historiae 7, I]} but would be a major step forward  to 
really build completely
predictive concrete string compactifications with
fluxes and intersecting D7-branes on such 
more involved Calabi-Yau orientifolds.


\section*{Acknowledgements}

We thank  N. Akerblom, M. Haack,
D. L\"ust, M. Schmidt-Sommerfeld and T. Weigand for interesting discussions.
We are also grateful to Volker Braun for sharing
part of his knowledge on algebraic geometry with us and
to J. Conlon and F. Quevedo for very useful and constructive comments
about the issues discussed  in this paper.

This work is supported in part by the European Community's Human
Potential Programme under contract MRTN-CT-2004-005104
`Constituents, fundamental forces and symmetries of the universe'.


\newpage
\appendix


\section{Cohomology classes of line bundles}
\label{app_cohom}

In this appendix we combinatorically compute the cohomology
classes of general line bundles ${\cal L}={\cal O}(m,n,l)$ over
the resolution ${\cal M}$ of the ambient space $\IP_{[1,3,3,3,5]}$.
The corresponding classes on the hypersurface ${\cal X}$ can then
be computed via the Koszul sequence.
\bea
\label{koszulb}
   0 \rightarrow {\cal L}\otimes {\cal O}(-15,-10,-5)_{\cal M} 
   \rightarrow {\cal L}_{\cal M}
        \rightarrow {\cal L}_{\cal X} \rightarrow 0 \;.
\eea
Let us recall the resolution
\eq{
  \renewcommand{\arraystretch}{1.1}
  \renewcommand{\arraycolsep}{7pt}
  \begin{array}{|c|c|c|c|c|c|c|}
  \hline
  x_1 & x_2 & x_3 & x_4 & x_5 & x_6 & x_7 \\ \hline\hline
  3 & 3 & 3 & 5 & 1 & 0 & 0 \\
  2 & 2 & 2 & 3 & 0 & 1 & 0 \\
  1 & 1 & 1 & 1 & 0 & 0 & 1 \\
  \hline
  \end{array}
}

\noindent 
Then the classes $H^i({\cal M},{\cal L})$ can be computed
by counting monomials of degree $(m,n,l)$ \cite{Distler:1996tj,Blumenhagen:1997vt} as listed in table \ref{monom}. 
This can be easily put on a computer. We have checked
for many examples that the results are consistent with the Euler characteristic
$\chi({\cal X},{\cal L})$ in eq. (\ref{euler}).

\begin{table}[ht]
\centering
\renewcommand{\arraystretch}{1.4}
\begin{tabular}{|c||c|}
\hline
  Cohomology & Monomials of degree $(m,n,l)$   \\ 
\hline\hline 
  $H^0({\cal M},{\cal L})$ & $P(x_1,x_2,x_3,x_4,x_5,x_6,x_7)$ \\[0.5mm]
\hline
  $H^1({\cal M},{\cal L})$ & ${P(x_1,x_2,x_3,x_6,x_7)\over 
    x_4 x_5 Q(x_4,x_5)}$ \ \  ${P(x_1,x_2,x_3,x_5,x_7)\over 
    x_4 x_6 Q(x_4,x_6)}$ \ \  ${P(x_1,x_2,x_3,x_4,x_6)\over 
    x_5 x_7 Q(x_5,x_7)}$\\
  & ${P(x_1,x_2,x_3,x_7)\over 
    x_4 x_5 x_6 Q(x_4,x_5,x_6)}$\ \  ${P(x_1,x_2,x_3,x_6)\over 
    x_4 x_5 x_7 Q(x_4,x_5,x_7)}$\\[0.7mm]
\hline
  $H^2({\cal M},{\cal L})$ & $0$ \\
\hline
  $H^3({\cal M},{\cal L})$ & ${P(x_4,x_5)\over  x_1 x_2 x_3 x_6 x_7
    Q(x_1,x_2,x_3,x_6,x_7) }$ \ \ ${P(x_4,x_6)\over  x_1 x_2 x_3 x_5
    x_7 Q(x_1,x_2,x_3,x_5,x_7) }$ \\
  & ${P(x_5,x_7)\over  x_1 x_2 x_3 x_4
    x_6 Q(x_1,x_2,x_3,x_4,x_6) }$ \ \  ${P(x_4,x_5,x_6)\over  
    x_1 x_2 x_3 x_7 Q(x_1,x_2,x_3,x_7) }$\\ 
  & ${P(x_4,x_5,x_7)\over  x_1 x_2 x_3 
    x_6 Q(x_1,x_2,x_3,x_6) }$ \\[0.7mm]
\hline
  $H^4({\cal M},{\cal L})$ & ${1\over x_1 x_2 x_3 x_4 x_5 x_6 x_7
    Q(x_1,x_2,x_3,x_4,x_5,x_6,x_7)}$ \\[0.7mm]
\hline
\end{tabular}
\caption{Cohomology groups and corresponding monomials for
$\IP_{[1,3,3,3,5]}[15]$.\label{monom}}
\end{table}


\section{The $\IP_{[1,1,3,10,15]}[30]$ Calabi-Yau}
\label{app_cy2}

Here we will briefly summarise some properties of the Calabi-Yau 
$\IP_{[1,1,3,10,15]}[30]$ as another example of a swiss-cheese like manifold. 
It has five K\"ahler moduli out of which four are toric.
In the following we collect the toric data for the resolution
of the toric singularities.

\begin{itemize}

\item The manifold is specified by the resolution
\eq{
  \renewcommand{\arraystretch}{1.1}
  \renewcommand{\arraycolsep}{7pt}  
  \begin{array}{|c|c|c|c|c|c|c|c||c|}
  \hline
  x_1 & x_2 & x_3 & x_4 & x_5 & x_6 & x_7 & x_8 & p \\ \hline\hline
  15&10 & 3 & 1 & 1 & 0 & 0 & 0 & 30\\
  5 & 3 & 1 & 0 & 0 & 1 & 0 & 0 & 10 \\
  3 & 2 & 0 & 0 & 0 & 0 & 1 & 0 & 6 \\
  6 & 4 & 1 & 0 & 0 & 0 & 0 & 1 & 12 \\
  \hline
  \end{array}
}

\item The Stanley-Reisner ideal reads
\eq{
  SR = \{\:x_2x_7\:,\:& x_2x_8\:,\: x_3x_8\:,\: x_1x_3x_6\:,\: 
  x_1x_6x_7\:,\\
  & x_1x_6x_8\:,\: 
  x_2x_4x_5\:,\: x_3x_4x_5\:,\: x_4x_5x_7\: \} \;.
}

\item The triple triple intersection numbers in the basis $\eta_1=D_5$, $\eta_2=D_6$, $\eta_3=D_7$, $\eta_4=D_8$ are encoded in
\eq{
  I_3 = -\eta_1^3 + 18\eta_2^3 +& 8\eta_3^3 + 9\eta_4^3 +
  2\eta_1^2\eta_2 + \eta_1^2\eta_4 - 6\eta_1\eta_2^2 \\
  & - 2\eta_1\eta_3^2 + \eta_3^2\eta_4 - 3\eta_1\eta_4^2 - 
  3\eta_3\eta_4^2 + \eta_1\eta_3\eta_4 \;.
}

\item If one expands the K\"ahler form in the basis $\{\eta_1,\eta_2,\eta_3,\eta_4\}$ as
\eq{
  J = t_1 \bigl[\eta_1\bigr]+t_2 \bigl[\eta_2\bigr]+t_3 
  \bigl[\eta_3\bigr]+t_4 \bigl[\eta_4\bigr] \;,
}
then the volumes of the basis divisors are
\eq{
  \tau_1 &= \frac{1}{2}\: \Bigl( 
    -t_1^2+4t_1t_2-6t_2^2-2t_3^2+2\bigl(t_1+t_3\bigr)t_4 -
    3t_4^2 \Bigr) \;,\\
  \tau_2 &= \Bigl( t_1-3t_2 \Bigr)^2\;,\\
  \tau_3 &= \frac{1}{2}\: \Bigl( -2t_1+4t_3+3t_4 \Bigr)
    \Bigl( 2 t_3-t_4 \Bigr) \;,\\
  \tau_4 &= \frac{1}{2}\:\Bigl( t_1+t_3-3 t_4 \Bigr)^2 \;.
}

\item 
The K\"ahler cone is found by imposing $\int_{\mathcal{C}} J>0$ which gives the following conditions on the $\{t_i\}$ \footnote{We are indebted to Volker Braun
for sharing his knowledge and computer program on the computation of 
the K\"ahler cone with us.}
\eq{
  \label{ex2_Kcone1}
  3t_2+t_3-3t_4 >0 \;,\quad
  t_1-3t_2 >0 \;,\quad
  t_4-t_3 >0 \;,\quad
  -t_1+2t_2+t_4>0 \;.
}

\item Using these restrictions, the overall volume is expressed in terms of the four-cycle volumina as
\eq{
  \label{ex2_vol1}
  \V = \frac{\sqrt{2}}{45} \left(
  \bigl( 15\tau_1 +5\tau_2 + 3\tau_3 + 6\tau_4 \bigr)^{3/2}
  -\bigl( 3\tau_3 + \tau_4 \bigr)^{3/2}
  -\frac{5}{\sqrt{2}}\: \tau_2^{3/2}
  -5 \tau_4^{3/2}
  \right) .
}
From this we see that by making $\tau_1$ large while keeping the others small, we obtain a swiss-cheese like structure.

\item The Euler characteristic $\chi$ for the cycle $D=m\, \eta_1+n\, \eta_2+p\,\eta_3+q\,\eta_4$ is
\eq{
  \chi\bigl(\mathcal{X},\mathcal{O}_{D}\bigr)=& 
  -3mn^2 + {\textstyle \frac{3}{2}}q^3 + 3n^3 - 
  {\textstyle \frac{1}{6}}m^3 + {\textstyle \frac{1}{2}}p^2q  
  + mpq - {\textstyle \frac{3}{2}}pq^2 
  - mp^2 \\& - 
  {\textstyle \frac{3}{2}}mq^2 + {\textstyle \frac{4}{3}}p^3 + 
  {\textstyle \frac{1}{2}}m^2q + m^2n - n - 
  {\textstyle \frac{1}{3}}p  
  - {\textstyle \frac{1}{2}}q 
  + {\textstyle \frac{13}{6}}m 
}
where $\mathcal{X}$ stands for $\IP_{[1,1,3,10,15]}[30]$. The interesting combinations for the present setup are those with $\chi=1$ and $m=0$. Up to wrapping numbers $100$, these are
\eq{
  \label{ex2_rigid1}
  (m,n,p,q) = (0,0,0,1)\;,\;\;
            (0,0,1,0)\;,\;\;
            (0,0,1,1)\;.
}

\item It is more convenient to work in a diagonal basis which we define guided by the form of the volume \eqref{ex2_vol1}
\eq{
  \begin{array}{l@{\hspace{3pt}}c@{\hspace{3pt}}lcl@{\hspace{3pt}}
  c@{\hspace{3pt}}l}
  D_a &=& 15 D_1 + 5 D_2 + 3 D_3 + 6 D_4\;, &\qquad&
  D_b &=& 3 D_3 + D_4 \;, \\
  D_c &=& D_2 \;, &\qquad&
  D_d &=& D_4 \;.
  \end{array}
}
In this basis the total volume reads 
\eq{
  \V = \frac{\sqrt{2}}{45} \left(
  \tau_a^{3/2}
  - \tau_b^{3/2}
  -\frac{5}{\sqrt{2}}\: \tau_c^{3/2}
  -5 \tau_d^{3/2}
  \right) \;,
}
and the triple intersection numbers again diagonalise
\eq{
   I_3= 225\, D_a^3 + 225\, D_b^3 + 18\, D_c^3 + 9\, D_d^3\;.
}

\item Expanding also the K\"ahler form in this diagonal basis as $J=t_a
  [D_a]-t_b [D_b]-t_c [D_c]-t_d[D_d]$, we find that the K\"ahler cone is
  defined by
\eq{
  \label{ex2_Kcone2}
  5\:t_b>t_d>t_c>0\;,\qquad
  t_a>t_b+2\:t_c+t_d\;.
}

\newpage

\item We finally present a list of monomials to be counted in order
to determine the cohomology classes
$H^i({\cal M},{\cal L})$ on the ambient toric variety.
We use the shorthand notation $(1,2,4,5,7,8\vert 3,6)$
for all monomials of the form ${P(x_1,x_2,x_4,x_5,x_7,x_8)\over x_3 x_6
  Q(x_3,x_6)}$ and similarly for the others.

\begin{table}[ht]
\centering
\renewcommand{\arraystretch}{1.4}
\begin{tabular}{|c||c|}
\hline
  Cohomology & Monomials of degree $(m,n,p,q)$   \\ 
\hline\hline 
  $H^0({\cal M},{\cal L})$ & $(1,2,3,4,5,6,7,8\vert )$ \\[0.5mm]
\hline
  $H^1({\cal M},{\cal L})$ & $(1,2,4,5,7,8\vert 3,6)$\
  $(1,2,4,5,6,8 \vert 3,7)$\ $(1,2,3,4,5,7 \vert 6,8)$\\ 
  & $(1,2,4,5,8 \vert 3,6,7)$\ $(1,2,4,5,7 \vert 3,6,8)$ \\[0.5mm]
\hline
  $H^2({\cal M},{\cal L})$ & $(3,4,5,6,8\vert 1,2,7)$\
  $(3,4,5,6,7\vert 1,2,8)$\ $(2,3,4,5,8\vert 1,6,7)$\\
 & $(1,3,6,7,8\vert 2,4,5)$\ $(1,2,6,7,8\vert 3,4,5)$\
 $(1,2,3,6,7\vert 4,5,8)$\\
& $(1,2,7,8\vert 3,4,5,6)$\ $(1,2,6,7\vert 3,4,5,8)$\
  $(2,3,4,5\vert 1,6,7,8)$\\
&  $(2,4,5,8\vert 1,3,6,7)$\  $(1,2,3,7\vert 4,5,6,8)$\
     $(1,2,6,8\vert 3,4,5,7)$\\
& $(3,4,5,6\vert 1,2,7,8)$\ $(3,4,5,8\vert 1,2,6,7)$\
   $(1,6,7,8\vert 2,3,4,5)$\\
& $(1,3,6,7\vert 2,4,5,8)$\  $(4,5,6,8\vert 1,2,3,7)$\
   $(3,4,5,7\vert 1,2,6,8)$\\
 & $(1,2,7\vert 3,4,5,6,8)$\
  $(1,2,8 \vert 3,4,5,6,7)$\ $(1,6,7\vert 2,3,4,5,8)$\\
 & $(2,4,5\vert 1,3,6,7,8)$\ $(3,4,5\vert 1,2,6,7,8)$\
 $(4,5,8\vert 1,2,3,6,7)$\\[0.5mm]
\hline
  $H^3({\cal M},{\cal L})$ & $(3,6 \vert 1,2,4,5,7,8)$\
  $(3,7\vert 1,2,4,5,6,8)$\ $(6,8\vert 1,2,3,4,5,7)$\\ 
 &  $(3,6,7\vert 1,2,4,5,8)$\ $(3,6,8\vert 1,2,4,5,7)$ \\[0.5mm]
\hline
  $H^4({\cal M},{\cal L})$ & $(\vert 1,2,3,4,5,6,7,8)$ \\[0.7mm]
\hline
\end{tabular}
\caption{Cohomology groups and corresponding monomials for
$\IP_{[1,1,3,10,15]}[30]$.}
\end{table}

\end{itemize}


\clearpage
\nocite{*}
\bibliography{rev}
\bibliographystyle{utphys}


\end{document}